\documentclass[twocolumn,preprintnumbers,amsmath,amsfonts,amssymb,floatfix,aps,pra,superscriptaddress]{revtex4-1}

\usepackage[T1]{fontenc}
\usepackage{graphicx}%
\usepackage{multirow}%
\usepackage{amsmath,amssymb,amsfonts}%
\usepackage{amsthm}%
\usepackage{mathrsfs}%
\usepackage[title]{appendix}%
\usepackage{xcolor}%
\usepackage{textcomp}%
\usepackage{booktabs}%
\usepackage{algorithm}%
\usepackage{algorithmicx}%
\usepackage{algpseudocode}%
\usepackage{listings}
\usepackage{bm}
\usepackage{braket}  
\usepackage{tikz}
\usetikzlibrary{arrows.meta}
\usepackage[x11names]{xcolor}
\usepackage{multibib}

\definecolor{falured}{rgb}{0.5, 0.09, 0.09}
\definecolor{dukeblue}{rgb}{0.0, 0.0, 0.61}
\definecolor{orange(colorwheel)}{rgb}{1.0, 0.5, 0.0}
\definecolor{orange-red}{rgb}{1.0, 0.27, 0.0}
\definecolor{darkgreen}{rgb}{0.0, 0.2, 0.13}
\definecolor{navy}{RGB}{0,0,128}
\definecolor{royalblue}{RGB}{0,82,155}
\definecolor{crimson}{RGB}{200,40,60}
\definecolor{teal}{RGB}{20,140,140}
\definecolor{deepblue}{RGB}{31,78,121}
\definecolor{slategray}{RGB}{90,90,90}
\definecolor{warmgray}{RGB}{130,130,130}
\definecolor{purpleaccent}{RGB}{120,80,170}
\definecolor{ballblue}{rgb}{0.13, 0.67, 0.8}

\newcommand{\be}{\begin{equation}}
\newcommand{\ee}{\end{equation}}

\begin{document}

\title{Benchmarking the fermionic quasi-1D many-body problem}

\date{\today}

\author{E. Gradova}%
\email{ekaterina.gradova@phys.ens.fr}
\affiliation{Laboratoire de Physique de l’\'Ecole normale sup\'erieure, ENS, Universit\'e PSL, CNRS, Sorbonne Universit\'e, Universit\'e Paris Cit\'e, F-75005 Paris, France
}

\author{F. Chevy}%
\email{frederic.chevy@phys.ens.fr}
\affiliation{Laboratoire de Physique de l’\'Ecole normale sup\'erieure, ENS, Universit\'e PSL, CNRS, Sorbonne Universit\'e, Universit\'e Paris Cit\'e, F-75005 Paris, France
}
\affiliation{Institut Universitaire de France (IUF), 75005 Paris, France}

\begin{abstract}
We investigate the validity of effective one-dimensional models for quasi-1D fermionic systems by benchmarking a coupled-channel approach against the exact low-energy theory derived from the underlying three-dimensional problem in the weakly and strongly attractive limits.

We show that reproducing the exact two-body scattering amplitude is insufficient to construct the correct effective low-energy theory of quasi-1D fermions. In the weakly attractive regime, it does not capture the emergent three-body interaction induced by transverse excitations. In the strongly attractive regime, it yields an atom-dimer scattering length with an incorrect dependence on the three-dimensional scattering length. These results demonstrate the limitations of effective one-dimensional descriptions based solely on two-body physics and highlight the need to explicitly include few-body correlations in quasi-1D systems.
\end{abstract}

\maketitle
\section{Introduction}
Thanks to their unprecedented tunability and degree of control, ultracold atomic gases have emerged as a versatile platform for the exploration of quantum many-body physics. In particular, techniques such as Feshbach resonances and optical or magnetic confinement \cite{bloch2008many} allow precise control over interactions and trapping potentials, enabling the realization of strongly correlated quantum matter in a wide range of experimental settings. In particular, dimensionality is a key parameter that governs the behavior of quantum many-body systems. Indeed, as the spatial dimension is reduced, fluctuations are enhanced and can profoundly alter the nature of many-body states, leading to striking phenomena absent in three dimensions. A paradigmatic example is provided by the Mermin–Wagner theorem \cite{mermin1966absence}, which forbids the emergence of long-range order in one and two dimensions. Likewise, in one dimension, the quasiparticle paradigm breaks down, and the low-energy physics is instead governed by collective excitations. These are described by Tomonaga–Luttinger liquid theory, where spin and charge degrees of freedom decouple \cite{giamarchi2003quantum,Kinoshita:2006,He2020EmergenceFermions}.
 
 Experimentally, low-dimensional systems are realized by strongly confining the gas in one or two spatial directions \cite{cazalilla2011one, Hadzibabic2011Two-dimensionalPerspective}. When all relevant energy scales are smaller than the excitation gap of the tightly confined directions, the corresponding degrees of freedom are frozen, and the system can be described by an effective low-dimensional model. In the ultracold regime, where interactions are typically well described by short-range potentials (except for strongly magnetic atoms \cite{DipolarAtomsPfau, DipolarAtomsFerlaino}), this reduction suggests an effective description in terms of zero-range interactions in reduced dimensions, such as the Lieb–Liniger model for bosons \cite{LiebLiniger_original} or the Yang–Gaudin model for fermions \cite{Yang1967SomeInteraction,Gaudin1967UnInteraction}.

However, this dimensional reduction is not always straightforward. Virtual excitations of the transverse modes and multichannel scattering processes can significantly renormalize the effective interaction, particularly in the vicinity of the Feshbach resonances used to tune interactions experimentally \cite{chin2010feshbach, LowDFeshbachSala}. As a result, the effective one-dimensional description may deviate substantially from simple zero-range models. Understanding how such effects modify the low-dimensional effective field theory of strongly confined resonant gases remains therefore an open problem, with direct relevance to ongoing experiments \cite{Sobirey2022ObservingSuperfluids, CIRNagerl} and theory \cite{Kestner2006ConditionsTrap,Kestner2007EffectiveTrap,Chevy2023AchievingFermions, pricoupenko19three}.

In this article, we address this question in the context of one-dimensional geometries. We show how exact results obtained in both weakly and strongly attractive regimes can be combined to construct an effective field theory describing the low-dimensional dynamics of a strongly confined gas. This approach reveals that, similarly to nuclear systems \cite{Hammer2013Colloquium:Nuclei},   three-body processes play a central role in determining the effective low-dimensional behavior of the system beyond the standard two-body description.

\section{The quasi-1D problem}
\label{Sec:Olshanii}

 The scattering of two ultracold particles of mass $m$ confined in a harmonic quantum waveguide of transverse frequency $\omega_\perp$ was studied theoretically in \cite{olshanii1998atomic}. In this article, it was shown that the 1D scattering amplitude could be written as 
 \be
 f_k=\frac{1}{a_\perp/a_{3D}+\zeta(1/2,-E/2\hbar\omega_\perp)},
\label{Eq:ScatteringOlshanii}
 \ee
where $a_\perp=\sqrt{2\hbar/m\omega_\perp}$ is the size of the ground state of the transverse confinement, $a_{3\mathrm{D}}$ is the scattering length that describes collisions at low energy,  $\zeta(\alpha,z)$ is Hurwitz's zeta function \cite{abramowitz1948handbook}, and $E=\hbar^2 k^2/m$ is the incoming energy of the pair. 

In the low-energy regime, this scattering amplitude can be expanded as 
\begin{equation}
f_k=\frac{1}{1+ik a_{1\mathrm{D}}+...},
\label{Eq:DiracDelta1D}
\end{equation}
where $a_{1\mathrm{D}}=-(a^2_{\perp}/2a_{3\mathrm{D}})(1+\zeta(1/2)a_{3\mathrm{D}}/a_\perp)$. This behavior is identical to what one would expect for a purely 1D contact potential $V_{1\mathrm{D}}=g_{1\mathrm{D}}\delta (z)$ with $g_{1\mathrm{D}}=-2\hbar^2/(ma_{1\mathrm{D}})$. When $a_{3\mathrm{D}}\rightarrow 0$, we recover Born's approximation prediction $a_{1\mathrm{D}}\sim -a_\perp^2/a_{3\mathrm{D}}$, while at the so-called confinement induced resonance $a_{\perp}/a_{3\mathrm{D}}=-\zeta(1/2)$, the effective coupling constant $g_{1\mathrm{D}}$ diverges and changes sign. 

In addition to describing the scattering properties, the poles of $f_k$ give the energies of the bound states of the two-body problem. One can prove that  a bound state exists for any value of the 3D scattering length \cite{Bergeman2003Atom-atomResonance} and that the corresponding binding energy $E_\mathrm{B}$ is a solution of the equation
\be
\zeta\biggl(\frac{1}{2},-\frac{E_\mathrm{B}}{2\hbar\omega_\perp} \biggr)=-\frac{a_\perp}{a_{3\mathrm{D}}}.
\label{Eq:BoundStateIlhanii}
\ee
In the weakly attractive limit $a_{3\mathrm{D}}\rightarrow 0^-$, the solutions of this equation can be approximated by $E_\mathrm{B}\sim -\hbar^2/ma_{1\mathrm{D}}^2$, which corresponds to the prediction of the 1D Dirac potential for a negative coupling constant $g_{1\mathrm{D}}$. In the opposite limit $a_{3\mathrm{D}}\rightarrow 0^+$, the scattering amplitude $a_{1\mathrm{D}}$ appearing in Eq. (\ref{Eq:DiracDelta1D}) becomes negative. However, for a true 1D-Dirac potential, a bound state can exist only for positive scattering length. The existence of a bound state beyond the confinement induced resonance predicted by Eq. (\ref{Eq:BoundStateIlhanii}) thus proves that a simple contact potential cannot capture at the same time the scattering properties and the bound state of a quasi-1D system and that a more general Hamiltonian must be found. 

\section{Coupled-channel model}
\label{Sec:CoupledChannel}

Coupled-channel models provide a way to solve the inconsistency between the quasi-1D regime and the Dirac potential properties in the strongly attractive limit $a_{3\mathrm{D}}\rightarrow 0^+$ \cite{Kestner2007EffectiveTrap}. We consider here a gas of spin 1/2 fermions confined in a tight waveguide. In a coupled-channel model, we add an effective bosonic degree of freedom to the Hilbert space that describes the deeply bound dimers. The simplest 1D Hamiltonian that incorporates these degrees of freedom can be written as

\be
\hat H_{\mathrm{cc}}=\hat T+\hat U+\hat V,
\label{Eq:TwoChannelHamiltonian}
\ee
with

\begin{eqnarray}
\widehat T&=&\sum_k \varepsilon_k (\widehat a_k^\dagger \widehat a_k+\widehat b_k^\dagger \widehat b_k)+(E_0+ \varepsilon_k/2)\widehat c_k^\dagger \widehat c_k,\\
\widehat U&=&\frac{\tilde g_{1\mathrm{D}}}{L}\sum_{k,k',q}\widehat b^\dagger_{k'+q}\widehat a^\dagger_{k-q} \widehat a_{k'} \widehat b_{k}, \\
\widehat V&=&\frac{\Gamma}{\sqrt{L}}\sum_{k,k'}\left[\widehat a_k^\dagger\widehat b_{k'}^\dagger\widehat c_{k+k'} +  \widehat c_{k+k'}^\dagger \widehat b_{k'} \widehat a_k\right]
\label{Eq:TwoChannelHamilotnian3}.
\end{eqnarray}
Here, $k$ is the momentum along $z$, $\epsilon_k=k^2/2m$ is the kinetic energy of one fermion, and $\hat a_k$ and $\hat b_k$ denote the annihilation operators of the two fermionic spin species. Bosonic degrees of freedom are described by the annihilation operator $\widehat c_k$. This Hamiltonian is parametrized by three effective physical quantities: $E_0$, which corresponds to the bare energy of the dimer, $\tilde g_{1\mathrm{D}}$ is an effective 1D coupling constant characterizing the magnitude of the 2-body contact interaction, and $\Gamma$ describes the creation and annihilation of dimers. These three parameters are chosen to match the properties of the full two-body problem discussed in the previous section. 

For this, let's consider the scattering of two particles described by the Hamiltonian (\ref{Eq:TwoChannelHamiltonian}). The associated T-matrix may be calculated exactly and is given by 

\be      T = \cfrac{1}{L} \,\, \cfrac{1}{\cfrac{1}{\tilde{g}_{\mathrm{1D}} + \frac{\Gamma^2}{E+i0^{+} - E_0}} + i\sqrt{\cfrac{\mu}{2\hbar^2 E}}}.
\label{Eq:transmittion}
\ee
We thus impose two constraints: (i) the pole of $T$ should be given by $E_\mathrm{B}$ solution of Eq. (\ref{Eq:BoundStateIlhanii}); (ii) the expansion of the scattering amplitude should coincide with the equivalent expression for Eq. (\ref{Eq:ScatteringOlshanii}) expanded to third order in $k$ in order to match both the 1D scattering length and the effective range. This yields a set of three conditions

\begin{eqnarray}
        E_0 &=& \cfrac{\biggl( \sqrt{ | E_\mathrm{B} | } + g_{1\mathrm{D}} \sqrt{\cfrac{m}{4\hbar^2}}   \biggr) E_\mathrm{B}}{\sqrt{ |E_\mathrm{B}| } - \sqrt{ \cfrac{m}{4\hbar^2}} \biggl( 
        \cfrac{\zeta(3/2) a_{\perp}^3}{8 a_{\mathrm{1D}}^2} E_\mathrm{B} - g_{1\mathrm{D}} \biggr)}, \\
        \Gamma &=& |E_0| \sqrt{\cfrac{\zeta(3/2)}{8}  \cfrac{a_{\perp}^{3}}{a_{1\mathrm{D}}^2}}, \\
         \tilde g_{1\mathrm{D}} &=& g_{1\mathrm{D}}+\frac{\Gamma^2}{E_0}=g_{1\mathrm{D}} + \cfrac{\zeta(3/2)}{8}  \cfrac{a_{\perp}^{3}}{a_{1\mathrm{D}}^2}  E_0.
\end{eqnarray}

\begin{figure*}[t]
    \begin{minipage}{0.33\textwidth}
    \centering{\includegraphics[width=1\linewidth]{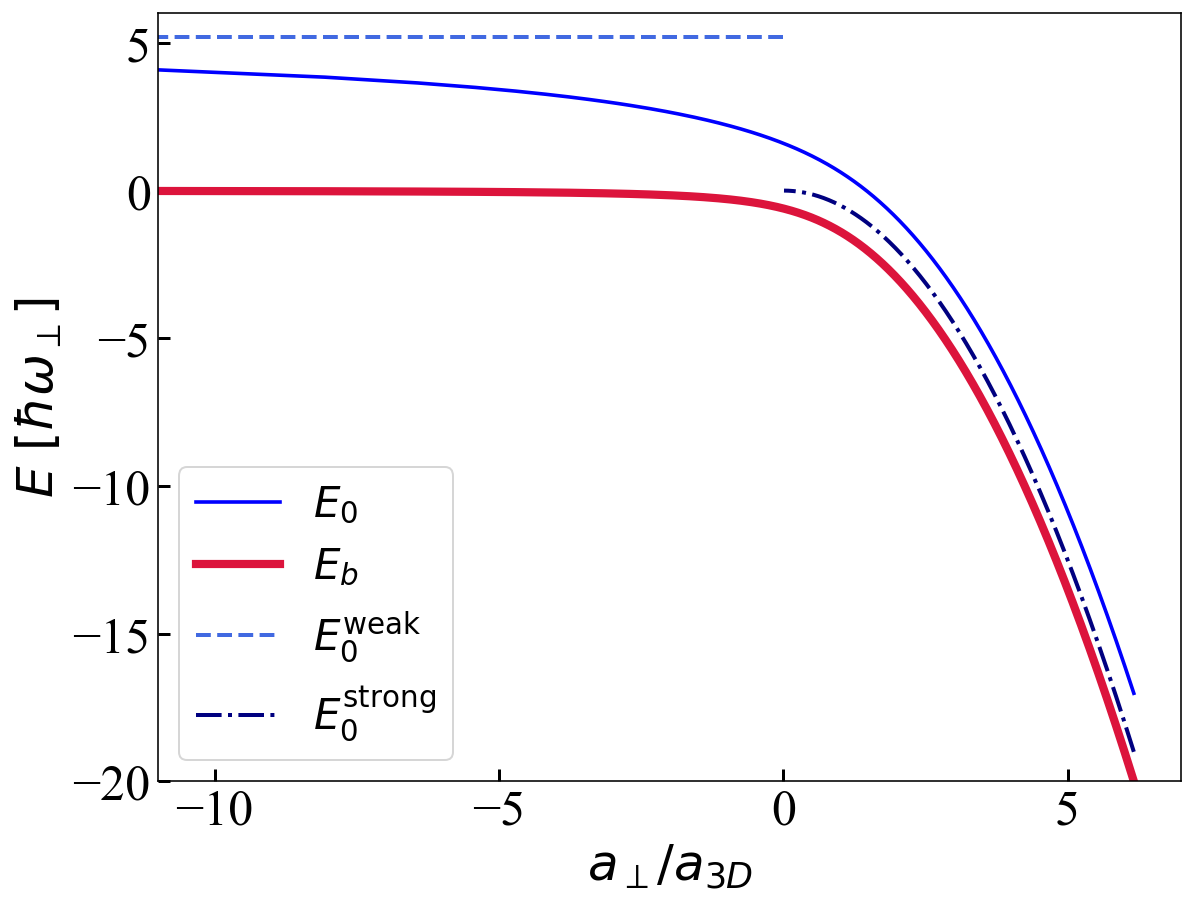}}
    \end{minipage}
    \hfill
    \begin{minipage}{0.33\textwidth}   \centering{\includegraphics[width=1\linewidth]{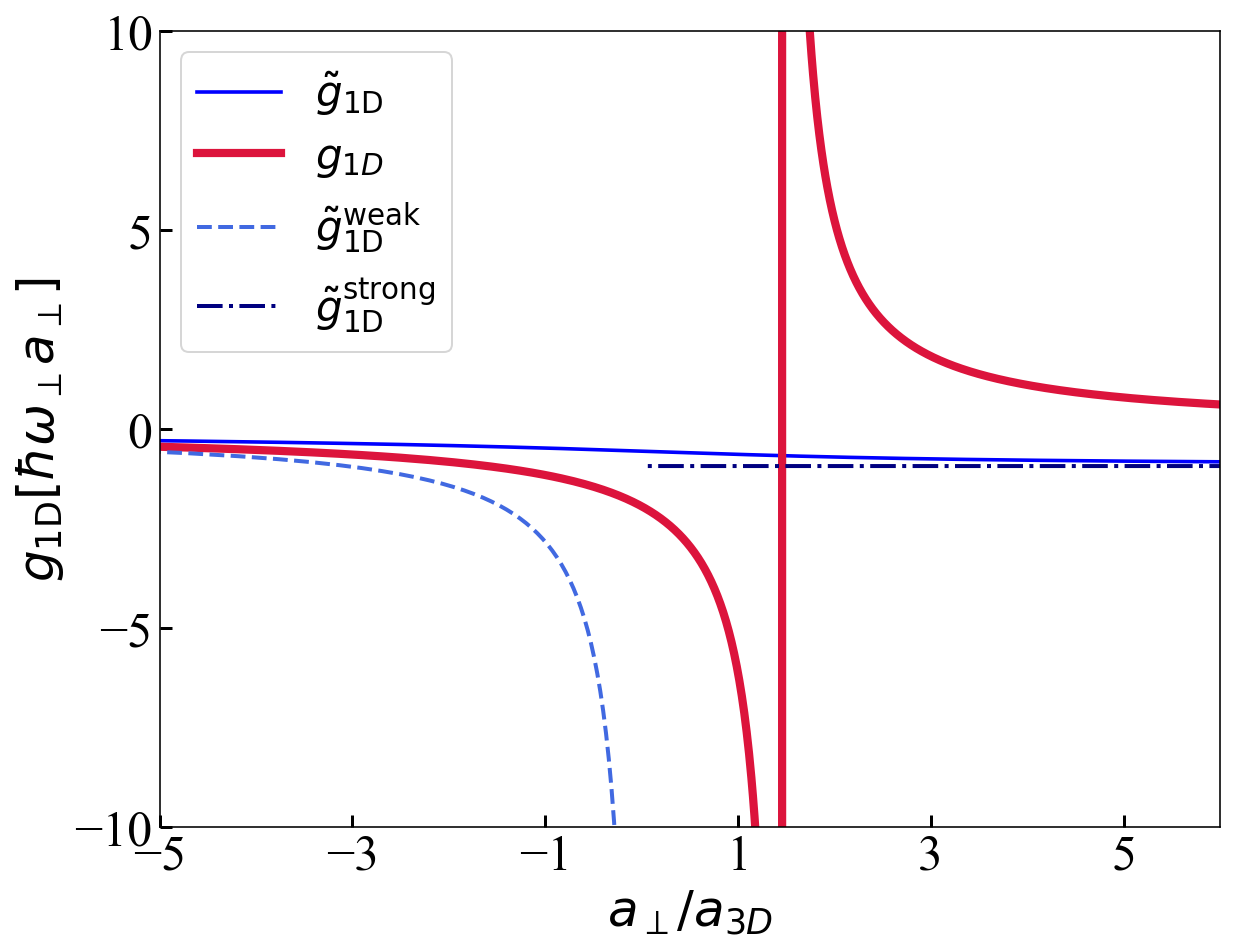}}
    \end{minipage}  
    \hfill
    \begin{minipage}{0.31\textwidth}
\centering{\includegraphics[width=1\linewidth]{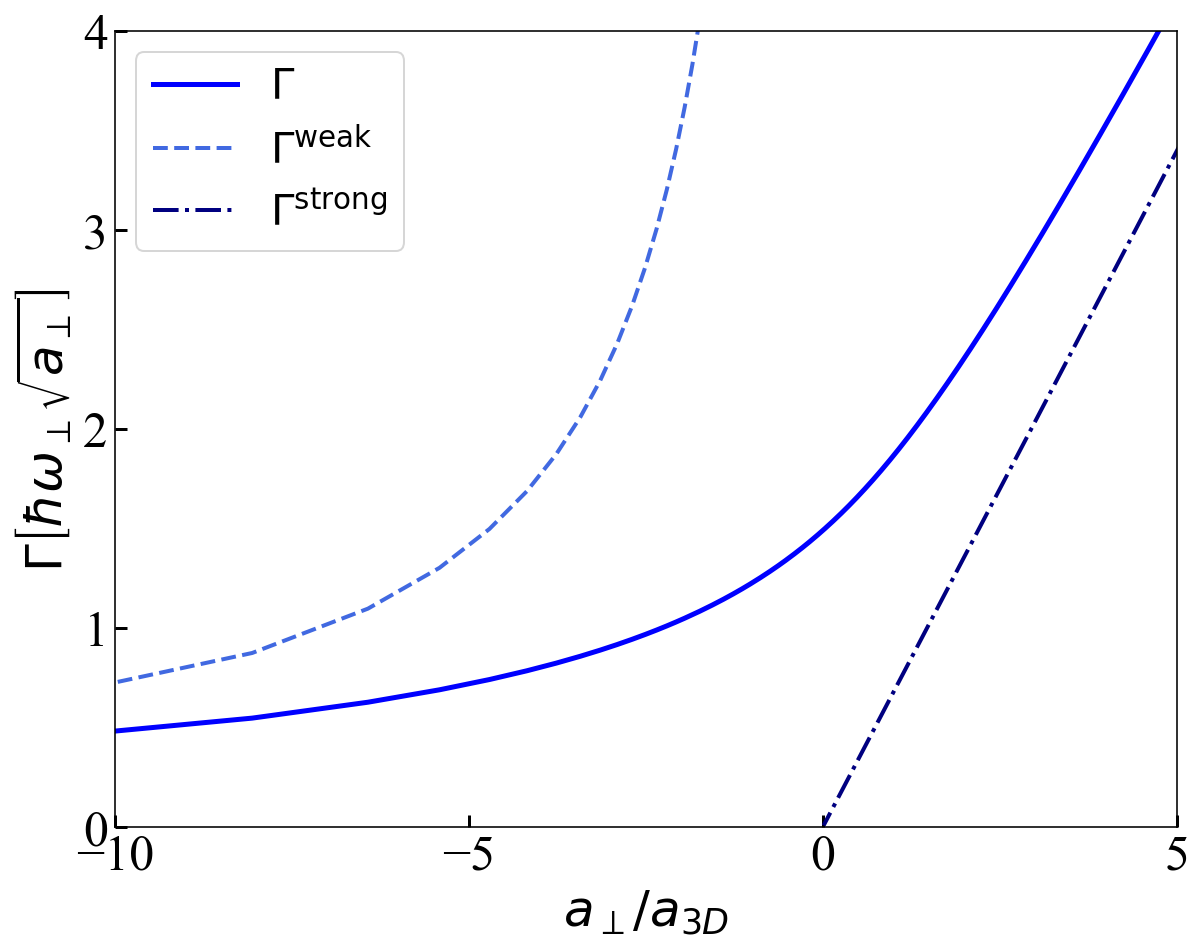}} 
    \end{minipage}

    \caption{Effective parameters of the coupled-channel Hamiltonian. From left to right: binding energy of bare molecules $E_0$ (with $E_\mathrm{B}$ given from equation \ref{Eq:BoundStateIlhanii}), coupling constant in 1D $g_{1\mathrm{D}}$ with coupling constant of contact interaction $\tilde g_{1\mathrm{D}}$, coupling constant of molecular channel $\Gamma$. The terms “weak” and “strong” denote expansion in the weakly and strongly attractive limits, respectively, given in the main text. The effective parameters in these limits are shown as dashed-dot dark blue and dashed light blue curves, respectively. }
    \label{Fig:EffectiveParameters}
\end{figure*}

The values of the different effective parameters are plotted as a function of the physical parameters $a_{3\mathrm{D}}$ and $a_\perp$ in Fig. (\ref{Fig:EffectiveParameters}). 


In the weakly attractive limit $a_{3\mathrm{D}}\rightarrow 0^-$, we obtain the following behaviors for the coupled-channel effective parameters  

\begin{eqnarray}
E_0&\simeq&\frac{8\zeta(3/2)}{3\zeta(5/2)}\hbar\omega_\perp\label{Eq:E0BCS},\\
\Gamma&\simeq&\frac{4\zeta(3/2)^{3/2}}{3\zeta(5/2)}\frac{a_{3\mathrm{D}}}{a_\perp}\hbar\omega_\perp\sqrt{2a_\perp}\label{Eq:Gamma0BCS},\\
\tilde g_{1\mathrm{D}}&\simeq&g_{1\mathrm{D}}\simeq 2\hbar\omega_\perp a_{3\mathrm{D}}\label{Eq:g1DBCS}.
\end{eqnarray}

In the strongly attractive limit $a_{3\mathrm{D}}\rightarrow 0^+$, the dressed and bare dimers coincide, and the effective parameters take the following forms:

\begin{eqnarray}
E_0&\simeq&E_\mathrm{B}\simeq -\frac{\hbar\omega_\perp}{2}\left(\frac{a_\perp}{a_{3\mathrm{D}}}\right)^2\label{Eq:E0BEC},\\
\Gamma&\simeq&\sqrt{\frac{\zeta(3/2)}{8}}\left(\frac{a_\perp}{a_{3\mathrm{D}}}\right)\hbar\omega_\perp\sqrt{a_\perp}\label{Eq:GammaBEC},\\
\tilde g_{1\mathrm{D}}&\simeq& -\frac{\zeta(3/2)}{4}\hbar\omega_\perp a_\perp\label{Eq:g1DBEC}.
\end{eqnarray}

We note that even though $g_{1\mathrm{D}}$ vanishes in this limit, the effective coupling $\tilde g_{1\mathrm{D}}$ converges to a constant, while the coupling $\Gamma$  diverges. The cancellation of $g_{1\mathrm{D}}$ therefore requires a fine tuning of the effective parameters $\tilde g_{1\mathrm{D}}$, $\Gamma$ and $E_0$ such that $\tilde g_{1\mathrm{D}}\simeq \Gamma^2/E_0$.

\section{Weakly attractive limit}

The coupled-channel model is designed to recover the exact low-energy physics in the two-body problem. We now explore its validity in more complex situations involving more than two particles, starting with the case of an ensemble of weakly attractive particles with $a_{3\mathrm{D}}\rightarrow 0^-$. In this case, it was recently shown that the true quasi-1D system could be described using an effective 1D Hamiltonian characterized by a two-body zero-range potential, plus some perturbative finite range corrections and an emerging three-body interaction \cite{Chevy2023AchievingFermions}. This approach is based on an elimination of the excited states of the transverse potential and provides an exact description of the low-energy behavior of the system. This effective quasi-1D Hamiltonian is then given by $\hat H_{\rm eff}=\hat H_1+\hat H_2+\hat H_3$, where $\hat H_n$ is the $n^{\mathrm{th}}$-body term and these contributions are, respectively, given by 
\begin{widetext}
\begin{eqnarray}
\hat H_1&=&\sum_k\varepsilon_k\left( \hat a^\dagger_k \hat a_k+\hat b^\dagger_k \hat b_k\right),\\
\hat H_2&=&\frac{1}{L}\sum_{\substack{k_1,k_2\\k'_1,k'_2\\k_1+k_2=k'_1+k'_2}}\left[g_{1\mathrm{D}}-\frac{\zeta(3/2)}{4}\frac{a_{3\mathrm{D}}^2}{a_\perp}\left(\epsilon_{k_1-k_2}/2+\epsilon_{k'_1-k'_2}/2\right)\right]\widehat a_{k_1}^\dagger \widehat b_{k_2}^\dagger \widehat b_{k'_2}\widehat a_{k'_1}
\label{Eq:Heff2},\\
\hat H_3&=&\frac{\tilde g_{3\mathrm{b}}}{L^2}\sum_{\substack{k_1+k_3+p_3\\=\\p_2+p_4+k_4}} 
\!\!\!\Big[(k_3-p_3)(k_4-p_4)
 \hat a_{k_1}^\dagger   \hat b_{k_3}^\dagger \hat b_{p_3}^\dagger b_{p_4} b_{k_4} a_{p_2}+ (\hat a \leftrightarrow \hat b)\Big],
\label{Eq:Heff3}\end{eqnarray}
\end{widetext}
where 
\be
\tilde g_{3\mathrm{b}}=-\frac{a_{3\mathrm{D}}^2}{4m}{\rm Li}_{1/2}(1/4),
\label{Eq:3bodycouplingconstant}
\ee
and $\mathrm{Li}_s(u)=\sum_{n=1}^\infty\frac{u^n}{n^s}$ is the so-called polylogarithm function.

$\hat H_1$ corresponds to the kinetic energy of the fermions. $\hat H_2$ is a two-body interaction made up of two contributions: the contact interaction proportional to the coupling constant $g_{1\mathrm{D}}$ found in the study of the two-body problem and a momentum-dependent term corresponding to finite range corrections. Finally, $\hat H_3$ describes an emergent three-body interaction resulting from virtual transitions to the transverse excited states.

By construction, the coupled-channel Hamiltonian is tuned to recover the properties of the two-body problem, and now we would like to test its validity in the many-body sector by comparing the action of $\hat H_{\rm cc}$ and $\hat H_{\rm eff}$ in the weakly attractive limit $a_{3D}\rightarrow 0^-$. To perform this comparison, we note that in this regime the binding energy of the bare molecules $E_0$ is of the order of $\hbar\omega_\perp$  (see Eq. \ref{Eq:E0BCS}) and is thus much larger than the typical kinetic energy in the quasi-1D limit $\varepsilon_k\ll \hbar\omega_\perp$. We can therefore eliminate the molecular degrees of freedom using  Schrieffer-Wolff's transformation  on the Hamiltonian $\hat H_{\rm cc}$ (\ref{Eq:TwoChannelHamiltonian}) \cite{Schrieffer1966RelationHamiltonians}.  More precisely, we perform the transformation on $\hat T$ and $\hat V$ only to cancel the action of $\hat V$ (which means that the action of the effective Hamiltonian will always be restricted to the dimer vacuum).

We thus introduce a unitary transformation $e^{\hat S}$ and define a new Hamiltonian

\be
\hat H'_{\rm cc}=e^{-\hat S}\hat H_{\rm cc} e^{\hat S}.
\ee
We choose $\hat S$ to eliminate $\hat V$ in the first order. $\hat S$ is thus proportional to $\Gamma$ and, as such, will be first order in $a_{3\mathrm{D}}$ (see Eq. (\ref{Eq:Gamma0BCS})). Expanding $\hat H'_{\rm cc}$ to the second order in $\hat S$ we have  

\be
\hat H'_{\rm cc}=\hat T+\hat U+\hat V+[\hat T,\hat S]+[\hat U,\hat S]+[\hat V,\hat S]+\frac{1}{2} [\hat S,[\hat S,\hat T]]+...
\ee
In order to eliminate $\hat V$, we chose $\hat S$ such that 

\be
\hat V+[\hat T,\hat S]=0.
\label{Eq:SW}
\ee
We thus have 

\be
\hat H'_{\rm cc}=\hat T+\hat U+[\hat U,\hat S]+\frac{1}{2} [\hat V,\hat S]+...
\ee
The following ansatz is assumed for $\hat S$:

\be
\hat S=\sum_{k_1,k_2}\left[\gamma_{k_1,k_2}\hat c_{k_1+k_2}^\dagger \hat a_{k_1}\hat b_{k_2}-\mathrm{ h.c.}\right].
\ee

Inserting this expression in Eq. (\ref{Eq:SW}), we get 

\begin{eqnarray}
\gamma_{k_1,k_2}&=&\frac{\Gamma}{\sqrt{L}}\frac{1}{E_0+\epsilon_{k_1+k_2}/2-\epsilon_{k_1}-\epsilon_{k_2}}\\
&=&\frac{\Gamma}{\sqrt{L}}\frac{1}{E_0-\epsilon_{k_1-k_2}/2}.
\end{eqnarray}

Using this expression, we see first that the commutator $[\hat U,\hat S]$ is a sum of terms proportional to $\hat c$ and  $\hat c^\dagger$. As a consequence, they will not contribute when projected on the dimer vacuum. The additional term contributing to the effective Hamiltonian is thus 

\be\frac{1}{2}[\widehat V,\widehat S]=
-\frac{1}{2}\sum_{\substack{k_1,k_2\\k'_1,k'_2\\k_1+k_2=k'_1+k'_2}}\left(\gamma_{k_1,k_2}+\gamma_{k'_1,k'_2}\right)\widehat a_{k_1}^\dagger \widehat b_{k_2}^\dagger \widehat b_{k'_2}\widehat a_{k'_1}.
\ee

In the quasi-1D limit, the kinetic energy is much smaller than $E_0\propto  \hbar\omega_\perp$ (See Eq. \ref{Eq:E0BCS}). Expanding $\gamma_{k_1k_2}$ to the leading order shows that the additional term $[\widehat V,\widehat S]/2$ is analogous to a contact interaction with a coupling constant $-\Gamma^2/E_0$. Combining this term with $\widehat U$ and using Eq. (\ref{Eq:g1DBCS}) recovers the coupling constant $g_{1\mathrm{D}}$ of the exact quasi-1D two-body problem \cite{olshanii1998atomic}. Defining $\widehat U'=\widehat U+[\widehat V,\widehat S]/2$, we thus have to the next-to-leading order

\be
\begin{split}
\widehat U'&=\frac{g_{1\mathrm{D}}}{L}\sum_{\substack{k_1,k_2\\k'_1,k'_2\\k_1+k_2=k'_1+k'_2}}\widehat a_{k_1}^\dagger \widehat b_{k_2}^\dagger \widehat b_{k'_2}\widehat a_{k'_1}-\\
&\frac{\zeta(3/2)}{4}\frac{a_{3\mathrm{D}}^2}{a_\perp L}\sum_{\substack{k_1,k_2\\k'_1,k'_2\\k_1+k_2=k'_1+k'_2}}\left(\epsilon_{k_1-k_2}/2+\epsilon_{k'_1-k'_2}/2\right)\widehat a_{k_1}^\dagger \widehat b_{k_2}^\dagger \widehat b_{k'_2}\widehat a_{k'_1}\\
&+...
\end{split}
\ee

We recover exactly the two-body term of $\hat H_{\rm eff}$ recalled in Eq. (\ref{Eq:Heff2}) and derived in  \cite{Chevy2023AchievingFermions} for the full quasi-1D system. This result is not surprising, since the coupled-channel model is fine-tuned to recover the exact two-body scattering amplitude at the effective-range level. 

However, our approach does not recover the three-body term Eq. (\ref{Eq:Heff3}), indicating that, beyond the two-body sector, the coupled-channel Hamiltonian $\hat H_{\rm cc}$ does not capture the full many-body physics. In order to recover exactly the  effective Hamiltonian $\hat H_{\rm eff}$, it is necessary to add a three-body contribution $\hat W$ to the Hamiltonian $\hat H_{\rm cc}$. At low energy, the identification with Eq. (\ref{Eq:Heff3}) yields 

\be
\begin{split}
\hat W&=\hat H_3\\
&=\frac{\tilde g_{3\rm b}}{L^2}\sum_{\substack{k_1+k_3+p_3\\=\\p_2+p_4+k_4}} 
\!\!\!\Big[(k_3-p_3)(k_4-p_4)
 \hat a_{k_1}^\dagger   \hat b_{k_3}^\dagger \hat b_{p_3}^\dagger b_{p_4} b_{k_4} a_{p_2}\\ &+ (\hat a \leftrightarrow \hat b)\Big].
\end{split}
\ee

\section{Strongly attractive limit: the three-body problem}

As mentioned above, the three-body physics  cannot be adequately captured in the weakly attractive limit using the coupled-channel model approach. In particular, to recover the correct second-order corrections beyond the mean-field limit, an explicit three-body interaction term should be included in the effective Hamiltonian describing the 1D-dynamics. We now turn to the strongly attractive limit, where we will show that these discrepancies become even more pronounced. In this limit, the deviations cannot be resolved within the framework of perturbative theory alone, but require non-perturbative corrections to the coupled-channel Hamiltonian.

Within the framework of the coupled-channel Hamiltonian, the three-body fermionic problem can be divided into two different sectors for the initial state: in the so-called "upper branch", we consider the scattering of three free fermions, while the "lower branch" describes the interaction between a molecule and a free fermion. The treatment of the upper branch is similar to that of the weakly attractive limit and gives rise to the emergent three-body interaction $\hat W$ obtained in the previous section and characterized by a coupling constant scaling as $a_{3\mathrm{D}}^2$ \cite{Chevy2023AchievingFermions}. In what follows, we therefore focus on the lower branch and study the scattering of an $ab$-molecule by a $b$-atom. 

In the strongly attractive limit, we have seen that the binding energy $E_0$ of the bare molecule was close to the physical binding energy $E_\mathrm{B}$. This suggests that in this regime the dimer wavefunction is mostly in the bare dimer state, and its projection on the continuum of scattering states is negligibly small. In order to clarify this point, we examine the Green's function of the molecule. Note $\hat G(\xi)=1/(\xi-\hat H_{\rm cc})$ the resolvent operator associated with the coupled-channel Hamiltonian $\hat H_{\rm cc}$ and $|\psi_0\rangle=\hat c_0|0\rangle$ the state that describes a bare dimer at rest. After a straightforward calculation, we obtain the following: 

\be
\langle \psi_0|\hat G (\xi)|\psi_0\rangle = \cfrac{1}{\xi-E_0+\cfrac{\Gamma^2}{\tilde g_{1D}+\sqrt{-2\xi\hbar^2 / m}}}.
\ee
Using Eq. (\ref{Eq:E0BCS},\ref{Eq:Gamma0BCS},\ref{Eq:g1DBCS}),  we see that the dimensionless numbers $\Gamma^2/{(\hbar^2\xi^3/m)^{1/2}}$ and $\tilde g_{1\mathrm{D}}/\sqrt{\hbar^2\xi/m}$ are vanishingly small close to the pole $\xi\simeq E_0$ and can therefore be neglected in the propagator. Physically, it means that the dressed dimer can be well described by a bare molecule.

\begin{figure}[h]
\centering
\begin{tikzpicture}[scale=0.7]
\draw[crimson, very thick, dash dot] (-4.5, 1.2) -- (4.5, 1.2);
\draw[deepblue, very thick] (-4.5, 0.6) -- (-1.3, 0.6);
\draw[slategray, thick] (1.3, 0.6) -- (4.5, 0.6);

\draw[slategray, thick] (-4.5, -1.5) -- (-1.3, -1.5);
\draw[deepblue, very thick] (1.3, -1.5) -- (4.5, -1.5);

\draw[deepblue, very thick] 
  (-1.3, 0.6) .. controls (-0.4, 0.6) and (0.1, -1.5) .. (1.3, -1.5);

\draw[slategray, thick] 
  (-1.3, -1.5) .. controls (-0.2, -1.5) and (0.4, 0.6) .. (1.3, 0.6);

\node[left] at (-4.75, 0.9) {$p$};
\node[right] at (4.75, 0.9) {$p'$};
\node[left] at (-4.75, -1.55) {$-p$};
\node[right] at (4.75, -1.55) {$-p'$};
\node[left] at (0.95, 1.65) {$p+p'$};

\fill[opacity=0] (-4.5,1.4) rectangle (1,2.2);
\draw[purpleaccent, thick, dashed] (-4.5,1.35) rectangle (-1.2,0.45);

\fill[opacity=0] (-4.5,1.4) rectangle (1,2.2);
\draw[purpleaccent, thick, dashed] (1.2,1.35) rectangle (4.5,0.45);

\node at (-4.7, 1.2) {$\downarrow$};
\node at (-4.7, 0.6) {$\uparrow$};
\node at (-4.7, -1.5) {$\uparrow$};
\end{tikzpicture}
\caption{The first-order three-body scattering diagram. The exchange of spin-down fermions occurs within the bare molecular state. The purple dashed rectangles represent a bare molecular state of two fermions with opposite spins}
\label{fig:first_order}
\end{figure}
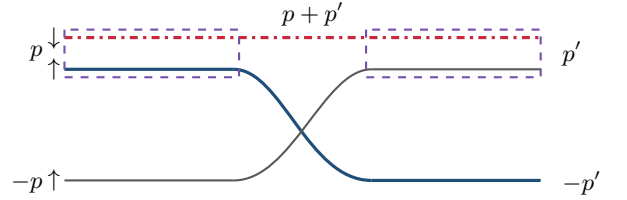

We therefore consider the $T$ matrix that describes the scattering from the initial state $|\psi(p)_i\rangle=\hat a_{-p}\hat c_p|0\rangle$ to the final state $|\psi_f(p')\rangle=\hat a_{-p'}\hat c_{p'}|0\rangle$, with $p'=\pm p$. To first order, the scattering is described by the diagram displayed in Fig. (\ref{fig:first_order})

 whose contribution is 

\be
T_{\mathrm{AD}} = \cfrac{\Gamma^2}{L}\frac{1}{E_0+3p^2/4m-(\epsilon_p+\epsilon_{p'}+\epsilon_{p+p'})+i0^+}.
 \ee

But, in the quasi-1D limit, we have $p^2/2m\ll \hbar\omega_\perp$, and in the strongly attractive limit, we further have $E_0\gg \hbar\omega_\perp$. We can therefore approximate the T-matrix as

\be
T_{\mathrm{AD}} \simeq \frac{\Gamma^2}{L E_0},
\ee
which would correspond to the first term of the expansion of the T-matrix associated with a contact atom-dimer potential $V_{\mathrm{AD}}=g^{\mathrm{AD}}_{1\mathrm{D}}\delta (z)$ with 
\be
g^{\rm AD}_{1D}\simeq \frac{\Gamma^2}{E_0}.
\ee
 
Using the asymptotic expression for $\Gamma$ and $E_0$ in the strongly attractive limit (Eq.  (\ref{Eq:E0BEC}) and (\ref{Eq:GammaBEC})), we have $a_{1\mathrm{D}}^{\mathrm{AD}} = 3 a_{\perp} / \zeta(3/2 ) $. 

This result can be compared to the exact calculation reported in \cite{Mora2004Atom-dimerGases}. For an atom and a fermion colliding in a harmonic waveguide, we indeed have the following
\begin{equation}
    a_{1 \mathrm{D} \, \mathrm{exact}}^{\mathrm{AD}} = - (a_{\perp}^{r})^2 / 2 (1.2 a_ {3 \mathrm{D}}),
\end{equation}
where $a_{\perp}^{r} = (3 \hbar / 2m \omega_{\perp})^{1/2}$ is the size of the ground state of the harmonic oscillator for the relative motion of the atom-dimer pair and $1.2 a_{3\mathrm{D}}$ corresponds to the 3D atom-dimer scattering length \cite{Mora2004Atom-dimerGases}. Comparing the two forms of the atom-dimer scattering length reveals that the coupled-channel model does not recover the correct scaling. In particular, the atom-dimer scattering length obtained within this model is independent of $a_{3\mathrm{D}}$, which is in direct contradiction to the exact three-body prediction. 

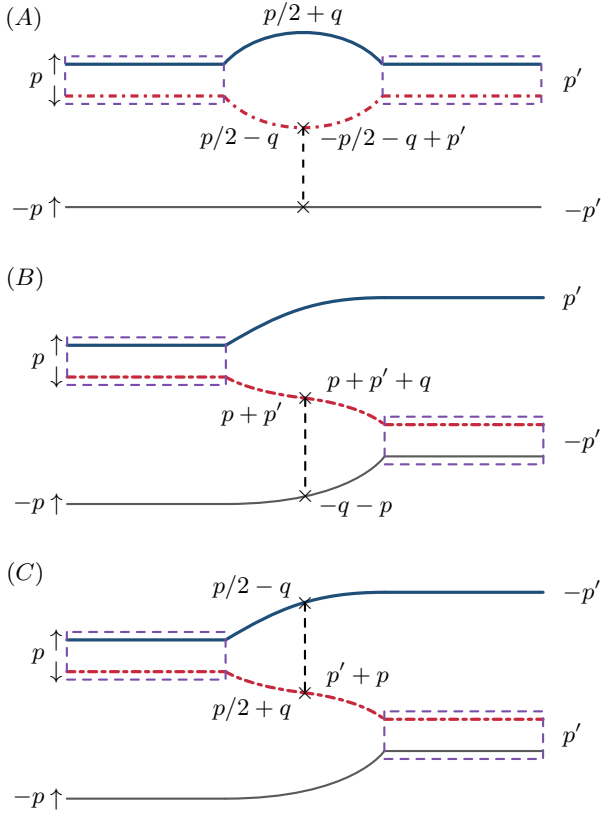
\begin{figure}

\begin{center}
\begin{tikzpicture}[thick, black, >=Stealth, scale=0.7]

\draw[deepblue, very thick] (-4.5, 1.2) -- (-1.5, 1.2);
\draw[deepblue, very thick] ( 1.5, 1.2) -- (4.5, 1.2);

\draw[crimson, very thick, dash dot] (-4.5, 0.6) -- (-1.5, 0.6);
\draw[crimson, very thick, dash dot] ( 1.5, 0.6) -- (4.5, 0.6);

\draw[slategray, thick] (-4.5, -1.5) -- (4.5, -1.5);

\draw[thick] 
  [deepblue, very thick](-1.5, 1.2) .. controls (-0.8, 2) and (0.8, 2) .. (1.5, 1.2);
\draw[thick][crimson, very thick, dash dot]
  (-1.5, 0.6) .. controls (-0.8, -0.2) and (0.8, -0.2) .. (1.5, 0.6);

\draw[thick] (0, 0) node[black] {$\times$};

\draw[thick] (0, -1.5) node[black] {$\times$};

\draw[black, dashed, thick] (0, 0) -- (0, -1.5);

\node[left] at (-4.75, 2.05) {$(A)$};

\node[left] at (-4.75, 0.9) {$p$};
\node[right] at (4.75, 0.9) {$p'$};

\node at (-4.7, 1.2) {$\uparrow$};
\node at (-4.7, 0.6) {$\downarrow$};
\node at (-4.7, -1.5) {$\uparrow$};

\node[left] at (-4.75, -1.55) {$-p$};
\node[right] at (4.75, -1.55) {$-p'$};

\node[below] at (0, 2.55) {$p/2+q$};

\node[below] at (-1.2, 0.2) {$p/2 - q$};   
\node[below] at (1.7, 0.2) {$-p/2 - q +p'$};

\fill[opacity=0] (-4.5,1.4) rectangle (1,2.2);
\draw[purpleaccent, thick, dashed] (-4.5,1.35) rectangle (-1.5,0.45);

\fill[opacity=0] (-4.5,1.4) rectangle (1,2.2);
\draw[purpleaccent, thick, dashed] (1.5,1.35) rectangle (4.5,0.45);

\end{tikzpicture}
\end{center}

\begin{center}
\begin{tikzpicture}[thick, black, line join=round, line cap=round, scale=0.7]

\draw[deepblue, very thick] (-4.5, 0.8) -- (-1.5, 0.8);
\draw[deepblue, very thick] (1.5, 1.7) -- (4.5, 1.7);

\draw[crimson, very thick, dash dot] (-4.5, 0.2) -- (-1.5, 0.2);

\draw[crimson, very thick, dash dot] (1.5, -0.7) -- (4.5, -0.7);
\draw[slategray, thick] (1.5, -1.3) -- (4.5, -1.3);

\draw[slategray, thick] (-4.5, -2.2) -- (-1.5, -2.2);

\draw[deepblue, very thick]
  (-1.5, 0.8) .. controls (-0.5, 1.4) and (0.2, 1.7) .. (1.5, 1.7);

\draw[crimson, very thick, dash dot]
  (-1.5, 0.2) .. controls (-1, -0.1) and (0, -0.2) .. (0, -0.2);
\draw[crimson, very thick, dash dot]
  (0, -0.2) .. controls (0, -0.2) and (1, -0.3) .. (1.5, -0.7);

\draw[slategray, thick]
  (-1.5, -2.2) .. controls (0.9, -2.2) and (1.5, -1.3) .. (1.5, -1.3);

\draw[dashed, thick, black] (0, -0.2) -- (0, -1.95);

\draw[thick] (0, -0.2) node[black] {$\times$};
\draw[thick] (0, -2.05) node[black] {$\times$};

\node[left] at (-4.75, 2.05) {$(B)$};

\node[left] at (-4.75, 0.5) {$p$};
\node[left] at (-4.75, -2.2) {$-p$};
\node[right] at (4.7, 1.7) {$p'$};
\node[right] at (4.7, -1) {$-p'$};

\node[below] at (-1, -0.1) {$p+p'$}; 
\node[above] at (1.4, -0.25) {$p+p'+q$}; 
\node[below] at (0.95, -1.9) {$-q-p$}; 

\node at (-4.7, 0.8) {$\uparrow$};
\node at (-4.7, 0.2) {$\downarrow$};
\node at (-4.7, -2.2) {$\uparrow$};

\fill[opacity=0] (-4,1.4) rectangle (1,2.2);
\draw[purpleaccent, thick, dashed] (-4.5,0.95) rectangle (-1.5, 0.05);

\fill[opacity=0] (-4,1.4) rectangle (1,2.2);
\draw[purpleaccent, thick, dashed] (1.5,-0.55) rectangle (4.5, -1.45);

\end{tikzpicture}
\end{center}

\begin{center}
\begin{tikzpicture}[thick, black, line join=round, line cap=round, scale=0.7]

\draw[deepblue, very thick] (-4.5, 0.8) -- (-1.5, 0.8);
\draw[deepblue, very thick] (1.5, 1.7) -- (4.5, 1.7);

\draw[crimson, very thick, dash dot] (-4.5, 0.2) -- (-1.5, 0.2);

\draw[crimson, very thick, dash dot] (1.5, -0.7) -- (4.5, -0.7);
\draw[slategray, thick] (1.5, -1.3) -- (4.5, -1.3);

\draw[slategray, thick] (-4.5, -2.2) -- (-1.5, -2.2);

\draw[deepblue, very thick]
  (-1.5, 0.8) .. controls (-0.5, 1.4) and (0.2, 1.7) .. (1.5, 1.7);

\draw[crimson, very thick, dash dot]
  (-1.5, 0.2) .. controls (-1, -0.1) and (0, -0.2) .. (0, -0.2);
\draw[crimson, very thick, dash dot]
  (0, -0.2) .. controls (0, -0.2) and (1, -0.3) .. (1.5, -0.7);

\draw[slategray, thick]
  (-1.5, -2.2) .. controls (0.9, -2.2) and (1.5, -1.3) .. (1.5, -1.3);

\draw[dashed, thick, black] (0, 1.5) -- (0, -0.2);

\draw[thick] (0, 1.5) node[black] {$\times$};
\draw[thick] (0, -0.2) node[black] {$\times$};

\node[left] at (-4.75, 2.05) {$(C)$};

\node[left] at (-4.75, 0.5) {$p$};
\node[left] at (-4.75, -2.2) {$-p$};
\node[right] at (4.7, 1.7) {$-p'$};
\node[right] at (4.7, -1) {$p'$};

\node at (-4.7, 0.8) {$\uparrow$};
\node at (-4.7, 0.2) {$\downarrow$};
\node at (-4.7, -2.2) {$\uparrow$};


\node[above] at (-1, 1.4) {$p/2-q$}; 
\node[below] at (-1, -0.1) {$p/2+q$}; 
\node[above] at (1, -0.25) {$p' + p$}; 

\fill[opacity=0] (-4,1.4) rectangle (1,2.2);
\draw[purpleaccent, thick, dashed] (-4.5,0.95) rectangle (-1.5, 0.05);

\fill[opacity=0] (-4,1.4) rectangle (1,2.2);
\draw[purpleaccent, thick, dashed] (1.5,-0.55) rectangle (4.5, -1.45);

\end{tikzpicture}
\end{center}

\caption{Diagrams representing the action of the contact interaction at leading order on atom-dimer scattering. Conventions are identical to the Fig \ref{fig:first_order}. The black vertical dashed lines indicate contact interactions between fermions.}
\label{fig:diagrams with contact interaction}
\end{figure}

This peculiar behavior arises from the divergence of $\Gamma$ in the strongly attractive regime  discussed at the end of Sec. (\ref{Sec:CoupledChannel}). Nevertheless, we saw that, in the case of the two-body problem, the contribution $\Gamma^2/E_0$ to the 1D coupling was exactly compensated by $\tilde g_{\mathrm{1D}}$, which eventually led to a vanishing coupling constant $g_{1\mathrm{D}}$ when $a_{\mathrm{3D}}\rightarrow 0^+$. In order to check whether this cancelation occurs also in the atom-dimer scattering, we note that $\tilde g_{1\mathrm{D}}$ appears at leading order in the class of diagrams shown in Fig. (\ref{fig:diagrams with contact interaction}). To evaluate this contribution, we sum over the complete set of these diagrams and obtain the following correction to the atom-dimer T-matrix

\be
\begin{split}
\delta T_{\mathrm{AD}} = &\cfrac{\tilde{g}_{\mathrm{1D}} \Gamma^2}{L^2} \sum_q \Biggl( \cfrac{1}{(\xi - E_{\mathrm{A}1}) (\xi - E_{\mathrm{A}2}) }\\ & - \cfrac{1}{(\xi - E_{\mathrm{B}1}) (\xi - E_{\mathrm{B}2}) } - \cfrac{1}{(\xi - E_{\mathrm{C}1}) (\xi - E_{\mathrm{C}2}) } \Biggr), 
\end{split}        
\ee

\begin{gather}
        E_{\mathrm{A}1} = E_{\mathrm{B}1} = \cfrac{\hbar^2}{2m} \biggl[ \biggl(\cfrac{p}{2} + q \biggr)^2 + \biggl(\cfrac{p}{2} -q \biggr)^2 + (-p)^2 \biggr],
        \\
        E_{\mathrm{A}2} = \cfrac{\hbar^2}{2m} \biggl[ \biggl(\cfrac{p}{2} + q \biggr)^2 + \biggl(-\cfrac{p}{2} -q + p' \biggr)^2 + (-p')^2 \biggr],
        \\ 
        E_{\mathrm{B}2} = E_{\mathrm{C}1} = \cfrac{\hbar^2}{2m} \biggl[ (-p')^2 + (p'+p)^2 + (-p)^2 \biggr],
        \\
        E_{\mathrm{C}2} = \cfrac{\hbar^2}{2m} \biggl[ (-p')^2 + (p+p'+q)^2 + (-q-p)^2 \biggr].
\end{gather}

Summing all three terms yields  $\delta T_{\mathrm{AD}} = 3 \zeta(3/2)^2 \hbar^2 \, a_{\mathrm{3D}} /16 m L a_{\perp}^2$, which vanishes for $a_{3\mathrm{D}}\rightarrow 0$ and  therefore will not be able to compensate for the leading order contribution, contrary to what was happening in the two-body sector.

We note that the three-fermion coupling $\hat W$ introduced earlier to provide a correct description of the dynamics in the upper branch scales as $a_{3\mathrm{D}}^2$. As a consequence, it cannot contribute to the cancellation of the atom-dimer scattering amplitude needed to reconcile the values of $a_{\mathrm{1D}}^{\mathrm{AD}}$ and $a_{\mathrm{1D \, exact}}^{\mathrm{AD}}$ \cite{Mora2004Atom-dimerGases}.

The discrepancy therefore calls for the addition of a new term to the effective Hamiltonian, and the only remaining possibility in the three-body sector is a direct atom-dimer coupling of the form

\begin{equation}
    \hat W' = \cfrac{\tilde g_{\mathrm{AD}}}{L} \sum_{pp'q} \left[\hat c^{+}_{p'+q} \hat a^{+}_{p-q} \hat a_{p} \hat c_{p'}+\hat c^{+}_{p'+q} \hat b^{+}_{p-q} \hat b_{p} \hat c_{p'}\right].
\label{Eq:AtomDimerCoupling}
\end{equation}

The value of the coupling constant $\tilde g_{\mathrm{AD}}$ is determined by treating $\hat W'$ within Born's approximation. By identifying the corresponding low-energy amplitude with the coupling constant associated with $a_{1 \mathrm{D} \, \mathrm{exact}}^{\mathrm{AD}}$, we readily find that $\tilde g_{\mathrm{AD}} = \hbar^2 / 2ma_{\perp} (\zeta(3/2)+
O(a_{3\mathrm{D}}))$ with the second term becoming dominant as $a_{3\mathrm{D}} \to 0$. 

The introduction of $\hat{W}'$ raises the question of its potential feedback on the upper-branch three-fermion amplitude, and conversely of the impact of $\hat{W}$ on the atom-dimer scattering length. As shown in Appendix~\ref{Sec:LowerUpper}, both cross-contributions are sub-leading and do not affect our conclusions.

\section{Conclusion}

In this work, we have benchmarked the validity of effective one-dimensional descriptions for quasi-1D fermionic systems by comparing a coupled-channel model to the exact low-energy theory derived from the full three-dimensional problem. Our analysis is restricted to the weakly and strongly attractive limits, where controlled approximations can be developed. 

In the weakly attractive regime, we have shown that the coupled-channel Hamiltonian correctly reproduces the two-body scattering properties, including effective-range corrections, but fails to capture the emergent three-body interaction arising from virtual excitations of the transverse modes. This demonstrates that, beyond the two-body sector, an effective description of quasi-1D systems necessarily requires the inclusion of explicit three-body terms.

In the strongly attractive regime, we have analyzed the atom-dimer scattering problem and found that the coupled-channel model predicts a non-zero atom-dimer scattering length that is independent of the three-dimensional scattering length. This result is in qualitative and quantitative disagreement with exact three-body calculations, which show that atom-dimer should vanish in this regime. This discrepancy highlights a fundamental limitation of the model: while it can be fine-tuned to reproduce two-body observables, it fails to capture the correct structure of three-body correlations.

Altogether, our results show that the breakdown of effective one-dimensional descriptions near confinement-induced resonances is not merely quantitative, but structural. In particular, both perturbative and non-perturbative regimes require the inclusion of genuine three-body processes to correctly describe the many-body physics. These findings suggest that the resonant regime, where no small parameter is available, will require from the outset a fully non-perturbative treatment incorporating few-body correlations that goes beyond the scope of the present work. 

More generally, our work provides a framework to assess and systematically improve effective low-dimensional Hamiltonians, opening the way to more accurate descriptions of strongly interacting quasi-1D quantum gases. 

Experimentally, these additional terms will manifest by a modification of the equation of state of the gas that can be quantified by a measure of the density profile of the cloud in the trap. Beyond thermodynamic signatures, another consequence is a breakdown of integrability that will add decay channels in the dynamics of the system and lead to an increased relaxation rate  \cite{Mazets08breakdown,Kristensen2016One-dimensionalIntegrability}.  

\appendix

\section{Full resummation of the atom-dimer T-matrix}
\label{Sec:TMatrixResummation}

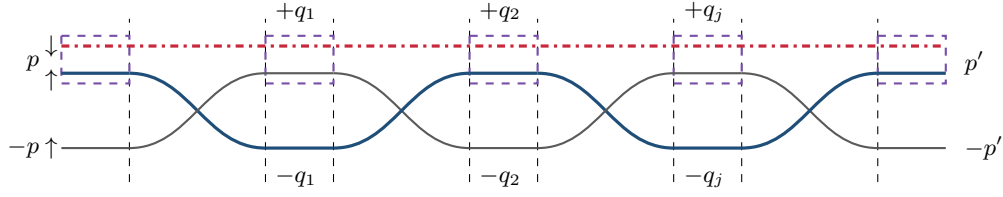
\begin{figure*}[t]
\centering
\begin{tikzpicture}[scale=0.9]
\node[align=center] at (3.45,2.5) {$+q_1$};
\node[align=center] at (3.45,0.07) {$-q_1$};
\node[align=center] at (6.45,2.5) {$+q_2$};
\node[align=center] at (6.45,0.07) {$-q_2$};
\node[align=center] at (9.45,2.5) {$+q_j$};
\node[align=center] at (9.45,0.07) {$-q_j$};

\node at (-0.15,2) {$\downarrow$};
\node at (-0.15,1.5) {$\uparrow$};
\node at (-0.15,0.5) {$\uparrow$};

\node[left] at (-0.2,1.75) {$p$};
\node[left] at (-0.2,0.5) {$-p$};
\node[right] at (13.15,1.75) {$p'$};
\node[right] at (13.15,0.5) {$-p'$};

\foreach \x in {1,3,4,6,7,9, 10, 12} {
    \draw[dashed] (\x,0) -- (\x,2.4);
}

\draw[crimson, very thick, dash dot] (0,2) -- (13,2);

\fill[opacity=0] (0,1.8) rectangle (1,2.2);
\draw[purpleaccent, thick, dashed] (0,2.15) rectangle (1,1.45);

\fill[opacity=0] (0,1.8) rectangle (1,2.2);
\draw[purpleaccent, thick, dashed] (3,2.15) rectangle (4,1.45);

\fill[opacity=0] (0,1.8) rectangle (1,2.2);
\draw[purpleaccent, thick, dashed] (6,2.15) rectangle (7,1.45);

\fill[opacity=0] (0,1.8) rectangle (1,2.2);
\draw[purpleaccent, thick, dashed] (9,2.15) rectangle (10,1.45);

\fill[opacity=0] (0,1.8) rectangle (1,2.2);
\draw[purpleaccent, thick, dashed] (12,2.15) rectangle (13,1.45);

\draw[deepblue, very thick] (0,1.6) -- (1,1.6);
\draw[slategray, thick] (0,0.5) -- (1,0.5);

\draw[deepblue, very thick]
    (1,1.6) to[out=0,in=180] (3,0.5);
\draw[slategray, thick]
    (1,0.5) to[out=0,in=180] (3,1.6);

\draw[slategray, thick] (3,1.6) -- (4,1.6);
\draw[deepblue, very thick] (3,0.5) -- (4,0.5);

\draw[slategray, thick]
    (4,1.6) to[out=0,in=180] (6,0.5);
\draw[deepblue, very thick]
    (4,0.5) to[out=0,in=180] (6,1.6);

\draw[deepblue, very thick] (6,1.6) -- (7,1.6);
\draw[slategray, thick] (6,0.5) -- (7,0.5);

\draw[deepblue, very thick]
    (7,1.6) to[out=0,in=180] (9,0.5);
\draw[slategray, thick]
    (7,0.5) to[out=0,in=180] (9,1.6);

\draw[slategray, thick] (9,1.6) -- (10,1.6);
\draw[deepblue, very thick] (9,0.5) -- (10,0.5);

\draw[slategray, thick]
    (10,1.6) to[out=0,in=180] (12,0.5);
\draw[deepblue, very thick]
    (10,0.5) to[out=0,in=180] (12,1.6);

\draw[deepblue, very thick] (12,1.6) -- (13,1.6);
\draw[slategray, thick] (12,0.5) -- (13,0.5);
\end{tikzpicture}
\caption{Three-body interaction diagram. The purple dashed rectangles represent a bare molecular state of two fermions with opposite spins. $q_j$ labels the intermediate momenta. }
\label{fig:3body_diagram_full}
\end{figure*}

Under the assumption that the dimer propagator is restricted to bare molecular states and that the effect of the contact interaction is negligible in the strongly attractive limit, we can actually resum the whole perturbation series describing atom-dimer scattering in the limit $p^2/2m\ll E_0$ for the atom-dimer interaction. Indeed, in this case, the only relevant diagrams are of the type in Fig. \ref{fig:3body_diagram_full}.

At order $n$, we have 

\begin{widetext}
\be
T_n=\left(\frac{\Gamma^2}{L}\right)^{n+1}\sum_{\{q_i\}} \left[\left(\prod_{i=1}^n\frac{1}{3p^4/4m-3q_i^2/4m+i0^+}\right)\left(\prod_{j=0}^n \frac{1}{E_0+3p^2/4m-\varepsilon_j(\left\{q_i\right\})}\right)\right],
\label{Eq:Tn}
\ee
\end{widetext}

 where $\varepsilon_j(\{q_i\})=q_j^2/m+q_{j+1}^2/m+q_j q_{j+1}/m$ (by convention, we take $q_0=p$ and $q_{n+1}=p'$). We observe that the first term (associated with bound atom pairs) is dominated by momenta $q_i\lesssim p$. By contrast, the terms corresponding to fully dissociated fermions are dominated by momenta $q_i\lesssim\sqrt{2m |E_0|}$. Since in the quasi-1D limit, we consider incoming momenta $p$ such that $p^2/2m\ll |E_0|$, we see that the latter condition is less restrictive than the former and we therefore conclude that we can approximate Eq. (\ref{Eq:Tn}) by 

\begin{eqnarray}
T_n&=&\left(\frac{\Gamma^2}{E_0L}\right)^{n+1}\left[ \sum_{q} \frac{1}{3p^4/4m-3q^2/4m+i0^+}\right]^n\\
&=&\frac{1}{L}\left(\frac{\Gamma^2}{E_0}\right)^{n+1}\left(\frac{2m}{3i\hbar p}\right)^n.
\label{Eq:Tn}
\end{eqnarray}
 The full T-matrix $T=\sum_n T_n$ is thus given by 

\be
T=\cfrac{1}{L}\, \cfrac{1}{E_0/\Gamma^2+2im / 3\hbar p},
\ee
 where we recognize the scattering amplitude for a 1D Dirac potential (Eq. \ref{Eq:DiracDelta1D}) associated with the expected coupling constant $g_{1\mathrm{D}}^{\mathrm AD}=2\hbar^2 / 3ma_{1\mathrm{D}}^{\rm AD}=\Gamma^2 / E_0$, 
 where the factor $2/3$ in the definition of the scattering length originates from the definition of the reduced mass of the atom-dimer system.  

\section{Lower branch versus upper branch in the strongly attractive limit}
\label{Sec:LowerUpper}

The conclusion of the article is that in the strongly attractive limit $a_{3\mathrm{D}}\rightarrow 0^+$, the effective 1D coupled-channel Hamiltonian takes the form 

\be
\hat H=\hat T+\hat U+\hat V+\hat W+\hat W',
\ee
where $\hat W$ corresponds to an emergent three-body interaction and is fixed by three-body scattering in the upper branch, while $\hat W'$ describes a contact interaction between a free fermion and a bare dimer and was tuned to recover the true quasi-1D atom-dimer scattering length in the lower branch.

One may wonder about a possible cross-talk between the lower and upper branches, in other words, whether $\hat W$ could have an impact on the atom dimer scattering length, or conversely whether $\hat W'$ could modify the three-fermions scattering amplitude. 

\begin{figure}
\begin{center}

\begin{tikzpicture}[thick, black, line join=round, line cap=round, scale=0.7]
\node[left] at (-5.6, 1.7) {$(i)$};
\draw[crimson, very thick, dash dot] (-4.5, 1.2) -- (-1.7, 1.2);
\draw[crimson, very thick, dash dot] (1.7, 1.2) -- (4.5, 1.2);

\draw[deepblue, very thick] (-4.5, 0.6) -- (-1.7, 0.6);
\draw[deepblue, very thick] (1.7, 0.6) -- (4.5, 0.6);

\draw[slategray, thick] (-4.5, -1.5) -- (-1.7, -1.5);
\draw[slategray, thick] (1.7, -1.5) -- (4.5, -1.5);

\draw[crimson, very thick, dash dot] 
  (-1.7, 1.2) .. controls (-0.8, 1.2) and (-0.2, 0.3) .. (-0.02, -0.55);

\draw[crimson, very thick, dash dot] 
  (-0.02, -0.55) .. controls (0.2, 0.3) and (0.8, 1.2) .. (1.7, 1.2);

\draw[deepblue, very thick] 
  (-1.7, 0.6) .. controls (-0.8, 0.6) and (-0.5, 0.5) .. (-0.02, -0.55);
\draw[slategray, thick] 
  (-1.7, -1.5) .. controls (-0.6, -1.5) and (-0.1, -1) .. (-0.02, -0.55);

\draw[slategray, thick] 
  (-0.02, -0.55) .. controls (0.1, -1) and (0.6, -1.5) .. (1.7, -1.5);

\draw[deepblue, very thick] 
  (-0.02, -0.55) .. controls (0.5, 0.5) and (0.8, 0.6) .. (1.7, 0.6);

\node[left] at (-4.75, 0.9) {$p$};
\node[right] at (4.75, 0.9) {$p'$};
\node[left] at (-4.75, -1.55) {$-p$};
\node[right] at (4.75, -1.55) {$-p'$};

\node[align=center] at (-1.25, 1.7) {$p/2 -q$};
\node[align=center] at (-1.4, -0.1) {$p/2 +q$};
\node[align=center] at (1.25, 1.7) {$p'/2 -q'$};
\node[align=center] at (1.4, -0.1) {$p'/2 +q'$};

\fill[opacity=0] (-4.5,1.4) rectangle (1,2.2);
\draw[purpleaccent, thick, dashed] (-4.5,1.35) rectangle (-1.7,0.45);

\fill[opacity=0] (-4.5,1.4) rectangle (1,2.2);
\draw[purpleaccent, thick, dashed] (1.7,1.35) rectangle (4.5,0.45);

\draw[thick] (-0.02, -0.55) node[black] {\Large $\times$};

\node at (-4.7, 1.2) {$\downarrow$};
\node at (-4.7, 0.6) {$\uparrow$};
\node at (-4.7, -1.5) {$\uparrow$};
\end{tikzpicture}

\begin{tikzpicture}[thick, black, line join=round, line cap=round, scale=0.7]
\node[left] at (-5.6, 2.8) {$(ii)$};

\draw[slategray, thick] (-4.5, 2.3) -- (4.5, 2.3);

\draw[crimson, very thick, dash dot] (-4.5, 0.9) -- (4.5, 0.9);

\draw[deepblue, very thick] 
  (-4.5, -0.9) .. controls (-2.4, -0.8) and (-1.3, 0.4) .. (-1.3, 0.4);
\draw[deepblue, very thick] (-1.3, 0.4) -- (1.3, 0.4);
\draw[deepblue, very thick] 
  (1.3, 0.4) .. controls (2.4, -0.8) and (4.5, -0.9) .. (4.5, -0.9);

\node[left] at (-4.75, 2.3) {$p_3$};
\node[right] at (4.75, 2.3) {$p'_3$};
\node[left] at (-4.75, 0.9) {$p_2$};
\node[right] at (4.75, 0.9) {$p'_2$};
\node[left] at (-4.75, -0.9) {$p_1$};
\node[right] at (4.75, -0.9) {$p'_1$};

\node at (-4.7, 2.3) {$\uparrow$};
\node at (-4.7, 0.9) {$\downarrow$};
\node at (-4.7, -0.9) {$\uparrow$};

\draw[black, dashed, thick] (0, 1.05) -- (0, 2.3);
\draw[thick] (0, 1.05) node[black] {$\times$};
\draw[thick] (0, 2.3) node[black] {$\times$};

\fill[opacity=0] (-4.5, 0.4) rectangle (4.5, 0.2);
\draw[purpleaccent, thick, dashed] (-1.3,1.05) rectangle (1.3,0.25);

\end{tikzpicture}
\end{center}
\caption{Lower (i) and upper (ii) branches coupling. Upper panel: impact of three-body emergent collisions on atom-dimer scattering. Lower panel: Impact of coupling between fermions and bare dimers on three-body scattering amplitude. }
\label{Fig:UpperLowerBranchCOupling}
\end{figure}
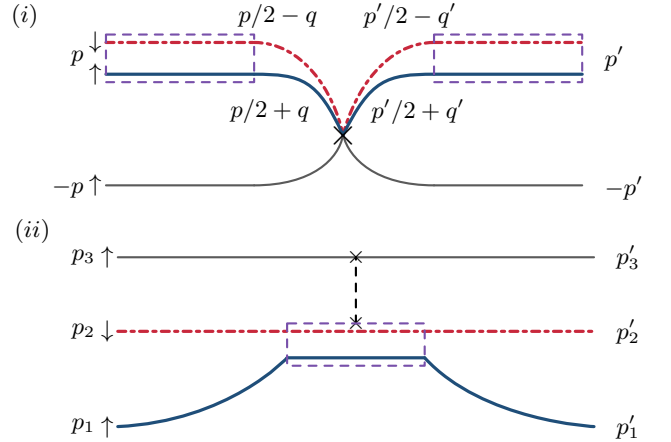

\begin{itemize}
\item[(i)] {\em Impact of $\hat W$ on the lower branch}. The modification of the atom-dimer T-matrix due to the three-body term of the Hamiltonian is given at the leading order by the diagram of Fig. \ref{Fig:UpperLowerBranchCOupling}(i), plus its permutations. All permutations give the same  contribution to the T-matrix, and we obtain 

\be
\delta T_{\mathrm{AD}}=4\frac{\tilde g_{3\mathrm{b}}\Gamma^2}{L^2}\sum_{q,q'}\frac{(3p/2+q)}{E_0-q^2/m}\frac{(3p'/2+q')}{E_0-{q'}^2/m}\propto \frac{\tilde g_{3\mathrm{b}}\Gamma^2}{E_0L}p^2.
\ee
We note first that this term is momentum dependent and thus corresponds to a finite range correction to the scattering potential. Second, using the scalings found for the parameters $\tilde g_{3\mathrm{b}}$, $\Gamma$ and $E_0$, we see that $\delta T_{\mathrm{AD}}\propto a_{3\mathrm{D}}^2$ and is thus not relevant at the order considered here (in Eq. (\ref{Eq:AtomDimerCoupling}), we neglect $O(a_{3\mathrm{D}})$).
\item[(ii)] {\em Impact of $\hat W'$ on the upper-branch}. The contribution of the atom-dimer interaction on the three-body scattering amplitude is given in the leading order by the diagrams shown in Fig. \ref{Fig:UpperLowerBranchCOupling}(ii). The correction to the three-body T-matrix is thus given by 
\begin{equation}
\begin{split}
    \delta T_{3\mathrm{b}}=& \cfrac{\Gamma^2 \tilde g_{\mathrm{AD}}}{L^2 (\epsilon_{p_1-p_2}/2 - E_0) (\epsilon_{p'_1-p'_2}/2 - E_0)}\\&\mathrm{+permutations}.
\end{split}
\end{equation}

Summing over all four permutations and expanding to leading order in momentum, we find that 

\be    
    \delta T_{3\mathrm{b}} = \frac{\Gamma^2 \tilde g_{\mathrm{AD}} }{16L^ 2E_0^4 m^2}(p_1 - p_3)(p'_1-p'_3)(p_1 + p_3 - 2p_2) (p'_1 + p'_3 - 2p'_2).
\ee

Using the expansion of the effective parameters with $a_{3\mathrm{D}}$, we see that this contribution scales as $a_{3\mathrm{D}}^6$, which is negligible compared to $\tilde g_{3\mathrm{b}}\propto a_{3\mathrm{D}}^2$.
\end{itemize}

The authors thank Xavier Leyronas for careful reading of the manuscript as well as all the members of LPENS ultracold many-body physics group for  insightful discussions. They acknowledge support from Institut Universitaire de France, PEPR Project Dyn-1D (ANR-23-PETQ-001), ANR Collaborative project ANR-25-CE47-6400 LowDCertif and ANR International Project ANR-24-CE97-0007 QSOFT.

\bibliography{references,new_references}

\begin{thebibliography}{28}%
\makeatletter
\providecommand \@ifxundefined [1]{%
 \@ifx{#1\undefined}
}%
\providecommand \@ifnum [1]{%
 \ifnum #1\expandafter \@firstoftwo
 \else \expandafter \@secondoftwo
 \fi
}%
\providecommand \@ifx [1]{%
 \ifx #1\expandafter \@firstoftwo
 \else \expandafter \@secondoftwo
 \fi
}%
\providecommand \natexlab [1]{#1}%
\providecommand \enquote  [1]{``#1''}%
\providecommand \bibnamefont  [1]{#1}%
\providecommand \bibfnamefont [1]{#1}%
\providecommand \citenamefont [1]{#1}%
\providecommand \href@noop [0]{\@secondoftwo}%
\providecommand \href [0]{\begingroup \@sanitize@url \@href}%
\providecommand \@href[1]{\@@startlink{#1}\@@href}%
\providecommand \@@href[1]{\endgroup#1\@@endlink}%
\providecommand \@sanitize@url [0]{\catcode `\\12\catcode `\$12\catcode
  `\&12\catcode `\#12\catcode `\^12\catcode `\_12\catcode `\%12\relax}%
\providecommand \@@startlink[1]{}%
\providecommand \@@endlink[0]{}%
\providecommand \url  [0]{\begingroup\@sanitize@url \@url }%
\providecommand \@url [1]{\endgroup\@href {#1}{\urlprefix }}%
\providecommand \urlprefix  [0]{URL }%
\providecommand \Eprint [0]{\href }%
\providecommand \doibase [0]{http://dx.doi.org/}%
\providecommand \selectlanguage [0]{\@gobble}%
\providecommand \bibinfo  [0]{\@secondoftwo}%
\providecommand \bibfield  [0]{\@secondoftwo}%
\providecommand \translation [1]{[#1]}%
\providecommand \BibitemOpen [0]{}%
\providecommand \bibitemStop [0]{}%
\providecommand \bibitemNoStop [0]{.\EOS\space}%
\providecommand \EOS [0]{\spacefactor3000\relax}%
\providecommand \BibitemShut  [1]{\csname bibitem#1\endcsname}%
\let\auto@bib@innerbib\@empty
\bibitem [{\citenamefont {Bloch}\ \emph {et~al.}(2008)\citenamefont {Bloch},
  \citenamefont {Dalibard},\ and\ \citenamefont {Zwerger}}]{bloch2008many}%
  \BibitemOpen
  \bibfield  {author} {\bibinfo {author} {\bibfnamefont {I.}~\bibnamefont
  {Bloch}}, \bibinfo {author} {\bibfnamefont {J.}~\bibnamefont {Dalibard}}, \
  and\ \bibinfo {author} {\bibfnamefont {W.}~\bibnamefont {Zwerger}},\
  }\href@noop {} {\bibfield  {journal} {\bibinfo  {journal} {Rev. Mod. Phys.}\
  }\textbf {\bibinfo {volume} {80}},\ \bibinfo {pages} {885} (\bibinfo {year}
  {2008})}\BibitemShut {NoStop}%
\bibitem [{\citenamefont {Mermin}\ and\ \citenamefont
  {Wagner}(1966)}]{mermin1966absence}%
  \BibitemOpen
  \bibfield  {author} {\bibinfo {author} {\bibfnamefont {N.~D.}\ \bibnamefont
  {Mermin}}\ and\ \bibinfo {author} {\bibfnamefont {H.}~\bibnamefont
  {Wagner}},\ }\href@noop {} {\bibfield  {journal} {\bibinfo  {journal}
  {Physical Review Letters}\ }\textbf {\bibinfo {volume} {17}},\ \bibinfo
  {pages} {1133} (\bibinfo {year} {1966})}\BibitemShut {NoStop}%
\bibitem [{\citenamefont {Giamarchi}(2003)}]{giamarchi2003quantum}%
  \BibitemOpen
  \bibfield  {author} {\bibinfo {author} {\bibfnamefont {T.}~\bibnamefont
  {Giamarchi}},\ }\href@noop {} {\emph {\bibinfo {title} {{Quantum physics in
  one dimension}}}},\ Vol.\ \bibinfo {volume} {121}\ (\bibinfo  {publisher}
  {Clarendon press},\ \bibinfo {address} {Oxford},\ \bibinfo {year}
  {2003})\BibitemShut {NoStop}%
\bibitem [{\citenamefont {Kinoshita}\ \emph {et~al.}(2006)\citenamefont
  {Kinoshita}, \citenamefont {Wenger},\ and\ \citenamefont
  {Weiss}}]{Kinoshita:2006}%
  \BibitemOpen
  \bibfield  {author} {\bibinfo {author} {\bibfnamefont {T.}~\bibnamefont
  {Kinoshita}}, \bibinfo {author} {\bibfnamefont {T.}~\bibnamefont {Wenger}}, \
  and\ \bibinfo {author} {\bibfnamefont {D.~S.}\ \bibnamefont {Weiss}},\
  }\href@noop {} {\bibfield  {journal} {\bibinfo  {journal} {Nature}\ }\textbf
  {\bibinfo {volume} {440}},\ \bibinfo {pages} {900} (\bibinfo {year}
  {2006})}\BibitemShut {NoStop}%
\bibitem [{\citenamefont {He}\ \emph {et~al.}(2020)\citenamefont {He},
  \citenamefont {Jiang}, \citenamefont {Lin}, \citenamefont {Hulet},
  \citenamefont {Pu},\ and\ \citenamefont {Guan}}]{He2020EmergenceFermions}%
  \BibitemOpen
  \bibfield  {author} {\bibinfo {author} {\bibfnamefont {F.}~\bibnamefont
  {He}}, \bibinfo {author} {\bibfnamefont {Y.~Z.}\ \bibnamefont {Jiang}},
  \bibinfo {author} {\bibfnamefont {H.~Q.}\ \bibnamefont {Lin}}, \bibinfo
  {author} {\bibfnamefont {R.~G.}\ \bibnamefont {Hulet}}, \bibinfo {author}
  {\bibfnamefont {H.}~\bibnamefont {Pu}}, \ and\ \bibinfo {author}
  {\bibfnamefont {X.~W.}\ \bibnamefont {Guan}},\ }\href {\doibase
  10.1103/PHYSREVLETT.125.190401/FIGURES/4/MEDIUM} {\bibfield  {journal}
  {\bibinfo  {journal} {Physical Review Letters}\ }\textbf {\bibinfo {volume}
  {125}},\ \bibinfo {pages} {190401} (\bibinfo {year} {2020})}\BibitemShut
  {NoStop}%
\bibitem [{\citenamefont {Cazalilla}\ \emph {et~al.}(2011)\citenamefont
  {Cazalilla}, \citenamefont {Citro}, \citenamefont {Giamarchi}, \citenamefont
  {Orignac},\ and\ \citenamefont {Rigol}}]{cazalilla2011one}%
  \BibitemOpen
  \bibfield  {author} {\bibinfo {author} {\bibfnamefont {M.~A.}\ \bibnamefont
  {Cazalilla}}, \bibinfo {author} {\bibfnamefont {R.}~\bibnamefont {Citro}},
  \bibinfo {author} {\bibfnamefont {T.}~\bibnamefont {Giamarchi}}, \bibinfo
  {author} {\bibfnamefont {E.}~\bibnamefont {Orignac}}, \ and\ \bibinfo
  {author} {\bibfnamefont {M.}~\bibnamefont {Rigol}},\ }\href@noop {}
  {\bibfield  {journal} {\bibinfo  {journal} {Reviews of Modern Physics}\
  }\textbf {\bibinfo {volume} {83}},\ \bibinfo {pages} {1405} (\bibinfo {year}
  {2011})}\BibitemShut {NoStop}%
\bibitem [{\citenamefont {Hadzibabic}\ and\ \citenamefont
  {Dalibard}(2011)}]{Hadzibabic2011Two-dimensionalPerspective}%
  \BibitemOpen
  \bibfield  {author} {\bibinfo {author} {\bibfnamefont {Z.}~\bibnamefont
  {Hadzibabic}}\ and\ \bibinfo {author} {\bibfnamefont {J.}~\bibnamefont
  {Dalibard}},\ }\href {\doibase 10.1393/NCR/I2011-10066-3} {\bibfield
  {journal} {\bibinfo  {journal} {La Rivista del Nuovo Cimento 2011 34:6}\
  }\textbf {\bibinfo {volume} {34}},\ \bibinfo {pages} {389} (\bibinfo {year}
  {2011})}\BibitemShut {NoStop}%
\bibitem [{\citenamefont {Lahaye}\ \emph {et~al.}(2009)\citenamefont {Lahaye},
  \citenamefont {Menotti}, \citenamefont {Santos}, \citenamefont {Lewenstein},\
  and\ \citenamefont {Pfau}}]{DipolarAtomsPfau}%
  \BibitemOpen
  \bibfield  {author} {\bibinfo {author} {\bibfnamefont {T.}~\bibnamefont
  {Lahaye}}, \bibinfo {author} {\bibfnamefont {C.}~\bibnamefont {Menotti}},
  \bibinfo {author} {\bibfnamefont {L.}~\bibnamefont {Santos}}, \bibinfo
  {author} {\bibfnamefont {M.}~\bibnamefont {Lewenstein}}, \ and\ \bibinfo
  {author} {\bibfnamefont {T.}~\bibnamefont {Pfau}},\ }\href {\doibase
  10.1088/0034-4885/72/12/126401} {\bibfield  {journal} {\bibinfo  {journal}
  {Reports on Progress in Physics}\ }\textbf {\bibinfo {volume} {72}},\
  \bibinfo {pages} {126401} (\bibinfo {year} {2009})}\BibitemShut {NoStop}%
\bibitem [{\citenamefont {Trautmann}\ \emph {et~al.}(2018)\citenamefont
  {Trautmann}, \citenamefont {Ilzhöfer}, \citenamefont {Durastante},
  \citenamefont {Politi}, \citenamefont {Sohmen}, \citenamefont {Mark},\ and\
  \citenamefont {Ferlaino}}]{DipolarAtomsFerlaino}%
  \BibitemOpen
  \bibfield  {author} {\bibinfo {author} {\bibfnamefont {A.}~\bibnamefont
  {Trautmann}}, \bibinfo {author} {\bibfnamefont {P.}~\bibnamefont
  {Ilzhöfer}}, \bibinfo {author} {\bibfnamefont {G.}~\bibnamefont
  {Durastante}}, \bibinfo {author} {\bibfnamefont {C.}~\bibnamefont {Politi}},
  \bibinfo {author} {\bibfnamefont {M.}~\bibnamefont {Sohmen}}, \bibinfo
  {author} {\bibfnamefont {M.~J.}\ \bibnamefont {Mark}}, \ and\ \bibinfo
  {author} {\bibfnamefont {F.}~\bibnamefont {Ferlaino}},\ }\href {\doibase
  10.1103/PhysRevLett.121.213601} {\bibfield  {journal} {\bibinfo  {journal}
  {Physical Review Letters}\ }\textbf {\bibinfo {volume} {121}},\ \bibinfo
  {pages} {213601} (\bibinfo {year} {2018})}\BibitemShut {NoStop}%
\bibitem [{\citenamefont {Lieb}\ and\ \citenamefont
  {Liniger}(1963)}]{LiebLiniger_original}%
  \BibitemOpen
  \bibfield  {author} {\bibinfo {author} {\bibfnamefont {E.~H.}\ \bibnamefont
  {Lieb}}\ and\ \bibinfo {author} {\bibfnamefont {W.}~\bibnamefont {Liniger}},\
  }\href {\doibase 10.1103/PhysRev.130.1605} {\bibfield  {journal} {\bibinfo
  {journal} {Physical Review}\ }\textbf {\bibinfo {volume} {130}},\ \bibinfo
  {pages} {1605} (\bibinfo {year} {1963})}\BibitemShut {NoStop}%
\bibitem [{\citenamefont {Yang}(1967)}]{Yang1967SomeInteraction}%
  \BibitemOpen
  \bibfield  {author} {\bibinfo {author} {\bibfnamefont {C.~N.}\ \bibnamefont
  {Yang}},\ }\href {\doibase 10.1103/PhysRevLett.19.1312} {\bibfield  {journal}
  {\bibinfo  {journal} {Physical Review Letters}\ }\textbf {\bibinfo {volume}
  {19}},\ \bibinfo {pages} {1312} (\bibinfo {year} {1967})}\BibitemShut
  {NoStop}%
\bibitem [{\citenamefont {Gaudin}(1967)}]{Gaudin1967UnInteraction}%
  \BibitemOpen
  \bibfield  {author} {\bibinfo {author} {\bibfnamefont {M.}~\bibnamefont
  {Gaudin}},\ }\href {\doibase 10.1016/0375-9601(67)90193-4} {\bibfield
  {journal} {\bibinfo  {journal} {Physics Letters A}\ }\textbf {\bibinfo
  {volume} {24}},\ \bibinfo {pages} {55} (\bibinfo {year} {1967})}\BibitemShut
  {NoStop}%
\bibitem [{\citenamefont {Chin}\ \emph {et~al.}(2010)\citenamefont {Chin},
  \citenamefont {Grimm}, \citenamefont {Julienne},\ and\ \citenamefont
  {Tiesinga}}]{chin2010feshbach}%
  \BibitemOpen
  \bibfield  {author} {\bibinfo {author} {\bibfnamefont {C.}~\bibnamefont
  {Chin}}, \bibinfo {author} {\bibfnamefont {R.}~\bibnamefont {Grimm}},
  \bibinfo {author} {\bibfnamefont {P.}~\bibnamefont {Julienne}}, \ and\
  \bibinfo {author} {\bibfnamefont {E.}~\bibnamefont {Tiesinga}},\ }\href@noop
  {} {\bibfield  {journal} {\bibinfo  {journal} {Reviews of Modern Physics}\
  }\textbf {\bibinfo {volume} {82}},\ \bibinfo {pages} {1225} (\bibinfo {year}
  {2010})}\BibitemShut {NoStop}%
\bibitem [{\citenamefont {Sala}\ \emph {et~al.}(2012)\citenamefont {Sala},
  \citenamefont {Schneider},\ and\ \citenamefont {Saenz}}]{LowDFeshbachSala}%
  \BibitemOpen
  \bibfield  {author} {\bibinfo {author} {\bibfnamefont {S.}~\bibnamefont
  {Sala}}, \bibinfo {author} {\bibfnamefont {P.~I.}\ \bibnamefont {Schneider}},
  \ and\ \bibinfo {author} {\bibfnamefont {A.}~\bibnamefont {Saenz}},\ }\href
  {\doibase 10.1103/PhysRevLett.109.073201} {\bibfield  {journal} {\bibinfo
  {journal} {Physical Review Letters}\ }\textbf {\bibinfo {volume} {109}},\
  \bibinfo {pages} {073201} (\bibinfo {year} {2012})}\BibitemShut {NoStop}%
\bibitem [{\citenamefont {Sobirey}\ \emph {et~al.}(2022)\citenamefont
  {Sobirey}, \citenamefont {Biss}, \citenamefont {Luick}, \citenamefont
  {Bohlen}, \citenamefont {Moritz},\ and\ \citenamefont
  {Lompe}}]{Sobirey2022ObservingSuperfluids}%
  \BibitemOpen
  \bibfield  {author} {\bibinfo {author} {\bibfnamefont {L.}~\bibnamefont
  {Sobirey}}, \bibinfo {author} {\bibfnamefont {H.}~\bibnamefont {Biss}},
  \bibinfo {author} {\bibfnamefont {N.}~\bibnamefont {Luick}}, \bibinfo
  {author} {\bibfnamefont {M.}~\bibnamefont {Bohlen}}, \bibinfo {author}
  {\bibfnamefont {H.}~\bibnamefont {Moritz}}, \ and\ \bibinfo {author}
  {\bibfnamefont {T.}~\bibnamefont {Lompe}},\ }\href {\doibase
  10.1103/PHYSREVLETT.129.083601/FIGURES/4/MEDIUM} {\bibfield  {journal}
  {\bibinfo  {journal} {Physical Review Letters}\ }\textbf {\bibinfo {volume}
  {129}},\ \bibinfo {pages} {083601} (\bibinfo {year} {2022})}\BibitemShut
  {NoStop}%
\bibitem [{\citenamefont {Haller}\ \emph {et~al.}(2010)\citenamefont {Haller},
  \citenamefont {Mark}, \citenamefont {Hart}, \citenamefont {Danzl},
  \citenamefont {Reichsöllner}, \citenamefont {Melezhik}, \citenamefont
  {Schmelcher},\ and\ \citenamefont {Nägerl}}]{CIRNagerl}%
  \BibitemOpen
  \bibfield  {author} {\bibinfo {author} {\bibfnamefont {E.}~\bibnamefont
  {Haller}}, \bibinfo {author} {\bibfnamefont {M.~J.}\ \bibnamefont {Mark}},
  \bibinfo {author} {\bibfnamefont {R.}~\bibnamefont {Hart}}, \bibinfo {author}
  {\bibfnamefont {J.~G.}\ \bibnamefont {Danzl}}, \bibinfo {author}
  {\bibfnamefont {L.}~\bibnamefont {Reichsöllner}}, \bibinfo {author}
  {\bibfnamefont {V.}~\bibnamefont {Melezhik}}, \bibinfo {author}
  {\bibfnamefont {P.}~\bibnamefont {Schmelcher}}, \ and\ \bibinfo {author}
  {\bibfnamefont {H.-C.}\ \bibnamefont {Nägerl}},\ }\href {\doibase
  10.1103/PhysRevLett.104.153203} {\bibfield  {journal} {\bibinfo  {journal}
  {Physical Review Letters}\ }\textbf {\bibinfo {volume} {104}},\ \bibinfo
  {pages} {153203} (\bibinfo {year} {2010})}\BibitemShut {NoStop}%
\bibitem [{\citenamefont {Kestner}\ and\ \citenamefont
  {Duan}(2006)}]{Kestner2006ConditionsTrap}%
  \BibitemOpen
  \bibfield  {author} {\bibinfo {author} {\bibfnamefont {J.~P.}\ \bibnamefont
  {Kestner}}\ and\ \bibinfo {author} {\bibfnamefont {L.-M.}\ \bibnamefont
  {Duan}},\ }\href {\doibase 10.1103/PhysRevA.74.053606} {\bibfield  {journal}
  {\bibinfo  {journal} {Physical Review A}\ }\textbf {\bibinfo {volume} {74}},\
  \bibinfo {pages} {053606} (\bibinfo {year} {2006})}\BibitemShut {NoStop}%
\bibitem [{\citenamefont {Kestner}\ and\ \citenamefont
  {Duan}(2007)}]{Kestner2007EffectiveTrap}%
  \BibitemOpen
  \bibfield  {author} {\bibinfo {author} {\bibfnamefont {J.~P.}\ \bibnamefont
  {Kestner}}\ and\ \bibinfo {author} {\bibfnamefont {L.~M.}\ \bibnamefont
  {Duan}},\ }\href {\doibase 10.1103/PHYSREVA.76.063610/FIGURES/1/MEDIUM}
  {\bibfield  {journal} {\bibinfo  {journal} {Physical Review A - Atomic,
  Molecular, and Optical Physics}\ }\textbf {\bibinfo {volume} {76}},\ \bibinfo
  {pages} {063610} (\bibinfo {year} {2007})}\BibitemShut {NoStop}%
\bibitem [{\citenamefont {Chevy}\ and\ \citenamefont
  {Orso}(2023)}]{Chevy2023AchievingFermions}%
  \BibitemOpen
  \bibfield  {author} {\bibinfo {author} {\bibfnamefont {F.}~\bibnamefont
  {Chevy}}\ and\ \bibinfo {author} {\bibfnamefont {G.}~\bibnamefont {Orso}},\
  }\href {\doibase 10.1103/PHYSREVA.107.043317/FIGURES/2/MEDIUM} {\bibfield
  {journal} {\bibinfo  {journal} {Physical Review A}\ }\textbf {\bibinfo
  {volume} {107}},\ \bibinfo {pages} {043317} (\bibinfo {year}
  {2023})}\BibitemShut {NoStop}%
\bibitem [{\citenamefont {Pricoupenko}(2019)}]{pricoupenko19three}%
  \BibitemOpen
  \bibfield  {author} {\bibinfo {author} {\bibfnamefont {L.}~\bibnamefont
  {Pricoupenko}},\ }\href {\doibase 10.1103/PhysRevA.99.012711} {\bibfield
  {journal} {\bibinfo  {journal} {Phys. Rev. A}\ }\textbf {\bibinfo {volume}
  {99}},\ \bibinfo {pages} {012711} (\bibinfo {year} {2019})}\BibitemShut
  {NoStop}%
\bibitem [{\citenamefont {Hammer}\ \emph {et~al.}(2013)\citenamefont {Hammer},
  \citenamefont {Nogga},\ and\ \citenamefont
  {Schwenk}}]{Hammer2013Colloquium:Nuclei}%
  \BibitemOpen
  \bibfield  {author} {\bibinfo {author} {\bibfnamefont {H.~W.}\ \bibnamefont
  {Hammer}}, \bibinfo {author} {\bibfnamefont {A.}~\bibnamefont {Nogga}}, \
  and\ \bibinfo {author} {\bibfnamefont {A.}~\bibnamefont {Schwenk}},\ }\href
  {\doibase 10.1103/RevModPhys.85.197} {\bibfield  {journal} {\bibinfo
  {journal} {Reviews of Modern Physics}\ }\textbf {\bibinfo {volume} {85}},\
  \bibinfo {pages} {197} (\bibinfo {year} {2013})}\BibitemShut {NoStop}%
\bibitem [{\citenamefont {Olshanii}(1998)}]{olshanii1998atomic}%
  \BibitemOpen
  \bibfield  {author} {\bibinfo {author} {\bibfnamefont {M.}~\bibnamefont
  {Olshanii}},\ }\href@noop {} {\bibfield  {journal} {\bibinfo  {journal}
  {Physical Review Letters}\ }\textbf {\bibinfo {volume} {81}},\ \bibinfo
  {pages} {938} (\bibinfo {year} {1998})}\BibitemShut {NoStop}%
\bibitem [{\citenamefont {Abramowitz}\ and\ \citenamefont
  {Stegun}(1948)}]{abramowitz1948handbook}%
  \BibitemOpen
  \bibfield  {author} {\bibinfo {author} {\bibfnamefont {M.}~\bibnamefont
  {Abramowitz}}\ and\ \bibinfo {author} {\bibfnamefont {I.~A.}\ \bibnamefont
  {Stegun}},\ }\href@noop {} {\emph {\bibinfo {title} {Handbook of mathematical
  functions with formulas, graphs, and mathematical tables}}},\ Vol.~\bibinfo
  {volume} {55}\ (\bibinfo  {publisher} {US Government printing office},\
  \bibinfo {year} {1948})\BibitemShut {NoStop}%
\bibitem [{\citenamefont {Bergeman}\ \emph {et~al.}(2003)\citenamefont
  {Bergeman}, \citenamefont {Moore},\ and\ \citenamefont
  {Olshanii}}]{Bergeman2003Atom-atomResonance}%
  \BibitemOpen
  \bibfield  {author} {\bibinfo {author} {\bibfnamefont {T.}~\bibnamefont
  {Bergeman}}, \bibinfo {author} {\bibfnamefont {M.~G.}\ \bibnamefont {Moore}},
  \ and\ \bibinfo {author} {\bibfnamefont {M.}~\bibnamefont {Olshanii}},\
  }\href {\doibase 10.1103/PHYSREVLETT.91.163201/FIGURES/2/MEDIUM} {\bibfield
  {journal} {\bibinfo  {journal} {Physical Review Letters}\ }\textbf {\bibinfo
  {volume} {91}},\ \bibinfo {pages} {163201} (\bibinfo {year}
  {2003})}\BibitemShut {NoStop}%
\bibitem [{\citenamefont {Schrieffer}\ and\ \citenamefont
  {Wolff}(1966)}]{Schrieffer1966RelationHamiltonians}%
  \BibitemOpen
  \bibfield  {author} {\bibinfo {author} {\bibfnamefont {J.~R.}\ \bibnamefont
  {Schrieffer}}\ and\ \bibinfo {author} {\bibfnamefont {P.~A.}\ \bibnamefont
  {Wolff}},\ }\href {\doibase 10.1103/PhysRev.149.491} {\bibfield  {journal}
  {\bibinfo  {journal} {Physical Review}\ }\textbf {\bibinfo {volume} {149}},\
  \bibinfo {pages} {491} (\bibinfo {year} {1966})}\BibitemShut {NoStop}%
\bibitem [{\citenamefont {Mora}\ \emph {et~al.}(2004)\citenamefont {Mora},
  \citenamefont {Egger}, \citenamefont {Gogolin},\ and\ \citenamefont
  {Komnik}}]{Mora2004Atom-dimerGases}%
  \BibitemOpen
  \bibfield  {author} {\bibinfo {author} {\bibfnamefont {C.}~\bibnamefont
  {Mora}}, \bibinfo {author} {\bibfnamefont {R.}~\bibnamefont {Egger}},
  \bibinfo {author} {\bibfnamefont {A.~O.}\ \bibnamefont {Gogolin}}, \ and\
  \bibinfo {author} {\bibfnamefont {A.}~\bibnamefont {Komnik}},\ }\href
  {\doibase 10.1103/PHYSREVLETT.93.170403/FIGURES/2/THUMBNAIL} {\bibfield
  {journal} {\bibinfo  {journal} {Physical Review Letters}\ }\textbf {\bibinfo
  {volume} {93}},\ \bibinfo {pages} {170403} (\bibinfo {year}
  {2004})}\BibitemShut {NoStop}%
\bibitem [{\citenamefont {Mazets}\ \emph {et~al.}(2008)\citenamefont {Mazets},
  \citenamefont {Schumm},\ and\ \citenamefont
  {Schmiedmayer}}]{Mazets08breakdown}%
  \BibitemOpen
  \bibfield  {author} {\bibinfo {author} {\bibfnamefont {I.~E.}\ \bibnamefont
  {Mazets}}, \bibinfo {author} {\bibfnamefont {T.}~\bibnamefont {Schumm}}, \
  and\ \bibinfo {author} {\bibfnamefont {J.}~\bibnamefont {Schmiedmayer}},\
  }\href {\doibase 10.1103/PhysRevLett.100.210403} {\bibfield  {journal}
  {\bibinfo  {journal} {Phys. Rev. Lett.}\ }\textbf {\bibinfo {volume} {100}},\
  \bibinfo {pages} {210403} (\bibinfo {year} {2008})}\BibitemShut {NoStop}%
\bibitem [{\citenamefont {Kristensen}\ and\ \citenamefont
  {Pricoupenko}(2016)}]{Kristensen2016One-dimensionalIntegrability}%
  \BibitemOpen
  \bibfield  {author} {\bibinfo {author} {\bibfnamefont {T.}~\bibnamefont
  {Kristensen}}\ and\ \bibinfo {author} {\bibfnamefont {L.}~\bibnamefont
  {Pricoupenko}},\ }\href {\doibase 10.1103/PhysRevA.93.023629} {\bibfield
  {journal} {\bibinfo  {journal} {Physical Review A}\ }\textbf {\bibinfo
  {volume} {93}},\ \bibinfo {pages} {023629} (\bibinfo {year}
  {2016})}\BibitemShut {NoStop}%
\end{thebibliography}%

\end{document}